\documentclass[twocolumn,nature communications,superacriptdaddress]{revtex4}

\usepackage{graphicx}
\usepackage{dcolumn}
\usepackage{bm}
\usepackage{amssymb}
\usepackage{amsfonts}
\usepackage{subfig}
\usepackage{latexsym}
\usepackage{enumerate}
\usepackage[dvipdfmx]{color} 
\usepackage{amsmath}
\usepackage[dvipdfmx]{epsfig}
\usepackage{times}
\usepackage{caption}

\newcommand{\red}{\color{black}}
\newtheorem{theorem}{Theorem}

\newtheorem{corollary}{Corollary}

\begin{document}
\widetext
\title{Optimal nonlocal conversion of photonic four-partite entanglement from two Bell pairs in quantum networks}
\author{Yuki Takeuchi}
\email{takeuchi@qi.mp.es.osaka-u.ac.jp}
\affiliation{Graduate School of Engineering Science, Osaka University, Toyonaka, Osaka 560-8531, Japan}
\author{Nobuyuki Imoto}
\email{imoto@mp.es.osaka-u.ac.jp}
\affiliation{Graduate School of Engineering Science, Osaka University, Toyonaka, Osaka 560-8531, Japan}
\author{Toshiyuki Tashima}
\email{tashima.toshiyuki.5e@kyoto-u.ac.jp}
\affiliation{Department of Electronic Science and Engineering, Kyoto University, Kyoto Daigaku-Katsura, Nishikyo-ku, Kyoto 615-8510, Japan}

\begin{abstract}
We analyze optimal schemes and also propose some practical schemes for the nonlocal conversion from two shared Bell pairs to four-qubit entangled states in optical quantum networks. In the analysis, we consider two-qubit operations as nonlocal operations and minimize the number of access to ancillary qubits as possible.
First, we consider two-qubit unitary operations without using ancillary qubits and derive a necessary and sufficient condition for convertible states. Second, we consider nonlocal optical systems composed of passive linear optics and postselection. For the passive linear optical systems, we derive achievable upper bounds of success probabilities of the conversion in the case without ancillary qubits. We also compare the optimal success probabilities with those of previously proposed schemes. Finally, we discuss success probabilities of the conversion in the case with ancillary qubits.
\end{abstract}

\maketitle 
\section{Introduction}
Multipartite entanglement is an important resource of a quantum network that enables several multiparty quantum information processing tasks such as quantum key distribution (QKD)~\cite{[E91]}, quantum teleportation~\cite{[BBCJPW93],[YC06]}, and distributed quantum computing~\cite{[BR03],[TKM05],[DP06]}. For each of the multiparty quantum information processing tasks, different entangled resource states are required. Particularly, Bell pairs, graph states~\cite{[BR01],[RB01],[RBB03]}, hypergraph states~\cite{[RHBM13]}, $W$ states~\cite{[DVC00],[KBI00],[D01]}, and Greenberger-Horne-Zeilinger (GHZ) states~\cite{[GHZ89]} are used in a variety of protocols as follows: Bell pairs can be used for QKD~\cite{[E91]}; quantum teleportation of a general one-qubit state~\cite{[BBCJPW93]}; and blind quantum computing~\cite{[BFK09],[TFIYI16],[HPF15]}. Graph states and hypergraph states can be used for universal quantum computing~\cite{[RB01],[RBB03],[MM16]}. $W$ states can be used for leader election in anonymous quantum networks~\cite{[DP06]} and asymmetric telecloning~\cite{[SIGA05]}. GHZ states can be used for achieving consensus in distributed networks without classical postprocessing~\cite{[DP06]} and secure delegated classical computing~\cite{[DKK16]}. Recently, in addition to these, Dicke states~\cite{[D54]} and $\chi$ states~\cite{[YC06]} have also been studied. For example, Dicke states can be used for Grover's quantum search algorithm~\cite{[IILV10]} and certain quantum versions of classical games~\cite{[SOMI04],[OSI07]}. $\chi$ states can be used for achieving optimal violation of a Bell inequality~\cite{[WYKO07]} and quantum teleportation of a general two-qubit state~\cite{[YC06]}.

So far, several generation methods of multipartite entangled states have been proposed~\cite{[KSTSW07],[WRZ05],[TOYKI08],[TOYKI09],[ITYKI11]}. In quantum networks, we normally consider the state generation of various multipartite entangled states by local operations and classical communication (LOCC) because of the characteristics of quantum entanglement. This is fundamentally interesting. On the other hand, it is well known that the conversion between any multipartite entangled state cannot be achieved using LOCC, because GHZ states, $W$ states, graph states, Dicke states, and $\chi$ states come from inequivalent entanglement classes under LOCC in the four-qubit case~\cite{[VDMV02],[SKLWZW08]}. Recently, in order to circumvent such no-go result, the nonlocal conversion required as the realistic networks has been actively studied. For example, it has been shown that a four-qubit linear cluster state can be probabilistically converted to a four-qubit GHZ state, a four-qubit Dicke state, and two Bell pairs~\cite{[TTONKW16],[MSMMSJTOT17]} using a tunable polarization-dependent beam splitter (PDBS) as a nonlocal operation. In another work, a universal optimal gate for transforming Dicke states has been proposed in the case where some qubits can be accessed from one node~\cite{[KIOTYKI14]}.
Furthermore, a scheme to fuse three $W$ states, which requires an ancillary qubit and access to one qubit of each of the $W$ states, has also been proposed~\cite{[OBYATO14]}.
However, the utility of nonlocal operations for the quantum network has not yet been fully understood. It is still challenging to achieve entanglement generation efficiently in restricted situations, where the number of access to qubits is minimized.

In this paper, we analyze the optimal state conversion from two Bell pairs to four-qubit entangled states and also propose some practical conversion schemes using restricted nonlocal operations. Here, we focus on a situation where one node is close to another one while far from the other two nodes for shared two Bell pairs. We also assume that nonlocal operations can be performed on only two nodes being close to each other. First, in Sec.~\ref{II}, we consider general two-qubit unitary operations as nonlocal operations. We then derive a necessary and sufficient condition for convertible states in this case. Second, in Sec.~\ref{III}, we consider optical systems composed of passive linear optics and postselection as nonlocal operations.
In Sec.~\ref{A}, we define nonlocal operations using only passive linear optics and postselection for our conversion schemes.
Then in Sec.~\ref{B}, we consider the nonlocal conversion from two Bell pairs to well-known four-qubit entangled states such as linear cluster states, GHZ states, $W$ states, Dicke states, different two Bell pairs, and $\chi$ states using the definition in Sec.~\ref{A}. In particular, we derive optimal success probabilities of the nonlocal conversion and show how to achieve the optimal success probabilities in passive linear optical systems. In Sec.~\ref{C}, we show improvements of success probabilities of the conversion when we use ancillary qubits. In Sec.~\ref{add}, we compare our scheme with some previous protocols and show advantages of our scheme. Section~\ref{IV} is devoted to the conclusion. In Appendix {\red A}, we provide details of derivation of optimal success probabilities for our nonlocal conversion methods. {\red In Appendix B, we derive the success probability and the fidelity of an output state of a nonlocal converter given in Sec.~\ref{B} when the transmittance of PDBSs is deviated from an ideal value.}

\begin{figure}[t]
\begin{center}
\includegraphics[width=6cm, clip]{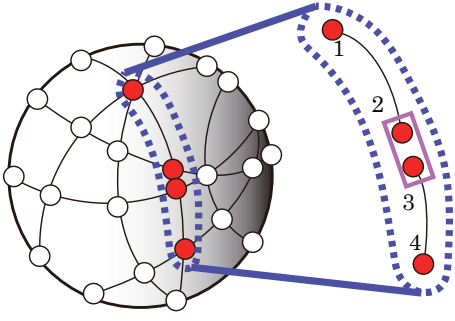}
\end{center}
\captionsetup{justification=raggedright,singlelinecheck=false}
\caption{An example of quantum networks. The image magnification is a four-qubit state shared by four nodes that are enclosed by dotted lines.}
\label{network}
\end{figure}

\medskip
\section{state conversion by nonlocal two-qubit unitary operations}
\label{II}
We focus on a shared four-qubit state in quantum networks, as shown in Fig.~\ref{network}. This four-qubit state is represented as two qubits close to each other (nodes $2$ and $3$) and other two qubits far from the others (nodes $1$ and $4$). In this situation, we consider the state conversion from two Bell pairs to four-qubit entangled states using two-qubit unitary operations performed on nodes $2$ and $3$. Two Bell pairs are given by $|\Phi^+\rangle_{1,2}|\Phi^+\rangle_{3,4}$, where $|\Phi^+\rangle_{i,j}\equiv(|HH\rangle_{i,j}+|VV\rangle_{i,j})/\sqrt{2}$. Here, $|H\rangle_i$ $(|V\rangle_i)$ represents a horizontally (vertically) polarized photon in node $i$ of the network. We define $V_{2,3}$ as a two-qubit unitary operator performed on nodes $2$ and $3$. We also define $U_1$ and $W_4$ as a single-qubit unitary operator performed on each node $1$ and $4$, respectively. Using these definitions and the initial state $|\Phi^+\rangle_{1,2}|\Phi^+\rangle_{3,4}$ with a relation given by \begin{eqnarray}
\label{bell}
U_j|\Phi^+\rangle_{i,j}=U_i^T|\Phi^+\rangle_{i,j},
\end{eqnarray} 
a converted four-qubit state $|f\rangle$ can be written as
\begin{eqnarray}
\nonumber
|f\rangle&\equiv& U_1V_{2,3}W_4|\Phi^+\rangle_{1,2}|\Phi^+\rangle_{3,4}\\
\nonumber
&=&V'_{2,3}|\Phi^+\rangle_{1,2}|\Phi^+\rangle_{3,4}\\
\label{SD}
&=&\cfrac{1}{2}\sum_{i,j\in\{H,V\}}|ij\rangle_{1,4}V'_{2,3}|ij\rangle_{2,3},
\end{eqnarray}
where $V'_{2,3}\equiv V_{2,3}U_2^TW_3^T$. From Eq.~(\ref{SD}), it is known that when $|f\rangle$ is divided into system A (the nodes $1$ and $4$) and system B (the nodes $2$ and $3$), the Schmidt rank of $|f\rangle$ is four, and all of Schmidt coefficients are equal to each other, i.e., $1/2$. It is a necessary condition for $|f\rangle$. Next, we show that this condition is also a sufficient condition for $|f\rangle$. Any four-qubit state that satisfy this condition can be written as
\begin{eqnarray*}
\cfrac{1}{2}\sum_{i,j=0}^{1}|\phi_{i,j}\rangle_{1,4}|\psi_{i,j}\rangle_{2,3}=\cfrac{1}{2}\sum_{i,j\in\{H,V\}}\tilde{V}_{1,4}|ij\rangle_{1,4}\tilde{V'}_{2,3}|ij\rangle_{2,3},
\end{eqnarray*}
where $\{|\phi_{i,j}\rangle|0\le i,j\le1\}$ and $\{|\psi_{i,j}\rangle|0\le i,j\le1\}$ are orthonormal bases of systems A and B, respectively. $\tilde{V}_{1,4}$ and $\tilde{V'}_{2,3}$ are two-qubit unitary operators on systems A and B, respectively. Moreover, any two-qubit unitary operator can be decomposed as
\begin{eqnarray}
\label{SU4}
(\tilde{U}\otimes\tilde{W}){\rm exp}[i(\theta_1X\otimes X+\theta_2Y\otimes Y+\theta_3Z\otimes Z)](\tilde{U'}\otimes\tilde{W'}),
\end{eqnarray}
where $\theta_i\in\mathbb{R}$ $(i=1,2,3)$~\cite{[ZVSW03]}. $\tilde{U}$, $\tilde{W}$, $\tilde{U'}$, and $\tilde{W'}$ are single-qubit unitary operators. $X$, $Y$, and $Z$ are Pauli $X$, $Y$, and $Z$ operators, respectively. For simplicity, we define $R\equiv{\rm exp}[i(\theta_1X\otimes X+\theta_2Y\otimes Y+\theta_3Z\otimes Z)]$. From Eqs.~(\ref{bell}) and (\ref{SU4}),
\begin{eqnarray}
\nonumber
&&\cfrac{1}{2}\sum_{i,j\in\{H,V\}}\tilde{V}_{1,4}|ij\rangle_{1,4}\tilde{V'}_{2,3}|ij\rangle_{2,3}\\
\label{U_14_23}
&=&\tilde{U}_1\tilde{W}_4R_{1,4}\tilde{U'}_1\tilde{W'}_4\tilde{V'}_{2,3}|\Phi^+\rangle_{1,2}|\Phi^+\rangle_{3,4}\\
\label{U_23}
&=&\tilde{V'}_{2,3}\tilde{U'}_2^T\tilde{W'}_3^TR_{2,3}\tilde{U}_2^T\tilde{W}_3^T|\Phi^+\rangle_{1,2}|\Phi^+\rangle_{3,4}\\
\nonumber
&=&|f\rangle.
\end{eqnarray}
This implies that the condition is also a sufficient condition for $|f\rangle$. As a result, the following theorem holds:
\begin{theorem}
\label{theorem1}
A quantum state can be converted from two Bell pairs $|\Phi^+\rangle_{1,2}|\Phi^+\rangle_{3,4}$ using two-qubit unitary operations performed on nodes $2$ and $3$ if and only if when the quantum state is divided into system A and system B, the Schmidt rank of the quantum state is four, and all of Schmidt coefficients are equal to $1/2$.
\end{theorem}
Note that in general, any two-qubit unitary operation is required for the conversion. Accordingly, this conversion scheme requires some non-linearities. For example, cross-Kerr non-linearity, an ancillary coherent state, and linear optics are sufficient to perform any two-qubit unitary operation on polarization-encoded qubits~\cite{[SNBMLM06]}.

Using Theorem~\ref{theorem1} and this situation, we examine the state conversion from two Bell pairs to well-known four-qubit entangled states. As well-known four-qubit entangled states, we focus on the linear cluster state $|C_4\rangle$, the GHZ state $|GHZ_4\rangle$, the $W$ state $|W_4\rangle$, the symmetric Dicke state $|D_4^{(2)}\rangle$, $|\Phi^+\rangle_{1,3}|\Phi^+\rangle_{2,4}$, $|\Phi^+\rangle_{1,4}|\Phi^+\rangle_{2,3}$, and the $\chi$ state $|\chi\rangle$ defined as follows:
\begin{widetext}
\begin{eqnarray}
\nonumber
|C_4\rangle&\equiv&\cfrac{|HHHH\rangle_{1,2,3,4}+|HHVV\rangle_{1,2,3,4}+|VVHH\rangle_{1,2,3,4}-|VVVV\rangle_{1,2,3,4}}{2}\\
\label{linear}
&=&\cfrac{|HH\rangle_{1,4}|HH\rangle_{2,3}+|HV\rangle_{1,4}|HV\rangle_{2,3}+|VH\rangle_{1,4}|VH\rangle_{2,3}-|VV\rangle_{1,4}|VV\rangle_{2,3}}{2}\\
\nonumber
|GHZ_4\rangle&\equiv&\cfrac{|HHHH\rangle_{1,2,3,4}+|VVVV\rangle_{1,2,3,4}}{\sqrt{2}}\\
\label{GHZ}
&=&\cfrac{|HH\rangle_{1,4}|HH\rangle_{2,3}+|VV\rangle_{1,4}|VV\rangle_{2,3}}{\sqrt{2}}\\
\nonumber
|W_4\rangle&\equiv&\cfrac{|HHHV\rangle_{1,2,3,4}+|HHVH\rangle_{1,2,3,4}+|HVHH\rangle_{1,2,3,4}+|VHHH\rangle_{1,2,3,4}}{2}\\
\label{W}
&=&\cfrac{1}{\sqrt{2}}\left(|HH\rangle_{1,4}\cfrac{|HV\rangle_{2,3}+|VH\rangle_{2,3}}{\sqrt{2}}+\cfrac{|HV\rangle_{1,4}+|VH\rangle_{1,4}}{\sqrt{2}}|HH\rangle_{2,3}\right)\\
\nonumber
|D_4^{(2)}\rangle&\equiv&\cfrac{1}{\sqrt{6}}(|HHVV\rangle_{1,2,3,4}+|HVHV\rangle_{1,2,3,4}+|VHHV\rangle_{1,2,3,4}+|HVVH\rangle_{1,2,3,4}+|VHVH\rangle_{1,2,3,4}\\
\nonumber
&&+|VVHH\rangle_{1,2,3,4})\\
\label{Dicke}
&=&\sqrt{\cfrac{2}{3}}\cfrac{|HV\rangle_{1,4}+|VH\rangle_{1,4}}{\sqrt{2}}\cfrac{|HV\rangle_{2,3}+|VH\rangle_{2,3}}{\sqrt{2}}+\cfrac{1}{\sqrt{6}}(|HH\rangle_{1,4}|VV\rangle_{2,3}+|VV\rangle_{1,4}|HH\rangle_{2,3})\\
\label{Bell_2}
|\Phi^+\rangle_{1,3}|\Phi^+\rangle_{2,4}&=&\cfrac{|HH\rangle_{1,4}|HH\rangle_{2,3}+|HV\rangle_{1,4}|VH\rangle_{2,3}+|VH\rangle_{1,4}|HV\rangle_{2,3}+|VV\rangle_{1,4}|VV\rangle_{2,3}}{2}\\
\label{Bell_3}
|\Phi^+\rangle_{1,4}|\Phi^+\rangle_{2,3}&=&\cfrac{|HH\rangle_{1,4}+|VV\rangle_{1,4}}{\sqrt{2}}\cfrac{|HH\rangle_{2,3}+|VV\rangle_{2,3}}{\sqrt{2}}\\
\nonumber
|\chi\rangle&\equiv&\cfrac{1}{2\sqrt{2}}(|HHHH\rangle_{1,2,3,4}-|HHVV\rangle_{1,2,3,4}-|HVHV\rangle_{1,2,3,4}+|VHHV\rangle_{1,2,3,4}+|HVVH\rangle_{1,2,3,4}\\
\nonumber
&&+|VHVH\rangle_{1,2,3,4}+|VVHH\rangle_{1,2,3,4}+|VVVV\rangle_{1,2,3,4})\\
\label{chi_2}
&=&\cfrac{1}{\sqrt{2}}\left(\cfrac{|HH\rangle_{1,4}+|VV\rangle_{1,4}}{\sqrt{2}}\cfrac{|HH\rangle_{2,3}+|VV\rangle_{2,3}}{\sqrt{2}}+\cfrac{|VH\rangle_{1,4}-|HV\rangle_{1,4}}{\sqrt{2}}\cfrac{|VH\rangle_{2,3}+|HV\rangle_{2,3}}{\sqrt{2}}\right)
\end{eqnarray}
\end{widetext}
From Eqs.~(\ref{linear})-(\ref{chi_2}) and Theorem~\ref{theorem1}, the following corollary holds:
\begin{corollary}
\label{corollary1}
Two Bell pairs $|\Phi^+\rangle_{1,2}|\Phi^+\rangle_{3,4}$ cannot be converted to $|GHZ_4\rangle$, $|W_4\rangle$, $|D_4^{(2)}\rangle$, $|\Phi^+\rangle_{1,4}|\Phi^+\rangle_{2,3}$, and $|\chi\rangle$ using two-qubit unitary operations performed on nodes $2$ and $3$. On the other hand, $|\Phi^+\rangle_{1,2}|\Phi^+\rangle_{3,4}$ can be converted to $|C_4\rangle$ and $|\Phi^+\rangle_{1,3}|\Phi^+\rangle_{2,4}$ using the same operations.
\end{corollary}
Note that even if we perform above nonlocal operations and two-qubit unitary operations on nodes $1$ and $4$, Theorem~\ref{theorem1} and Corollary~\ref{corollary1} hold. It is clear from Eqs.~(\ref{U_14_23}) and (\ref{U_23}).

\medskip
\section{Nonlocal conversion using passive linear optics and postselection}
\label{III}
Corollary~\ref{corollary1} means that $|\Phi^+\rangle_{1,2}|\Phi^+\rangle_{3,4}$ cannot be converted to any four-qubit state using only two-qubit unitary operations performed on nodes $2$ and $3$. In order to convert to more various classes of the entanglement composed of four-qubit states in optical quantum networks, we consider optical systems based on a postselection as nonlocal operations. The optical systems considered in this section are composed of passive linear optics, i.e., PDBSs, polarization-independent beam splitters, phase shifters (PSs), and wave plates. We also assume the same situation as considered in Sec.~\ref{II}. In this situation, we show that the passive linear optical systems enable us to convert from two Bell pairs to entangled states that cannot be converted in Sec.~\ref{II}. To this end, in Sec.~\ref{A}, we define nonlocal operations using only passive linear optics and postselection. In Sec.~\ref{B}, using the definition of the nonlocal operation given in Sec.~\ref{A}, we derive achievable upper bounds of success probabilities of the conversion from shared two Bell pairs to well-known four-qubit entangled states. We then show that existing optical systems can achieve the upper bounds. In Sec.~\ref{C}, we give improvements of the conversion when we use ancillary qubits.

\begin{figure*}[t]
\begin{center}
\includegraphics[width=10cm, clip]{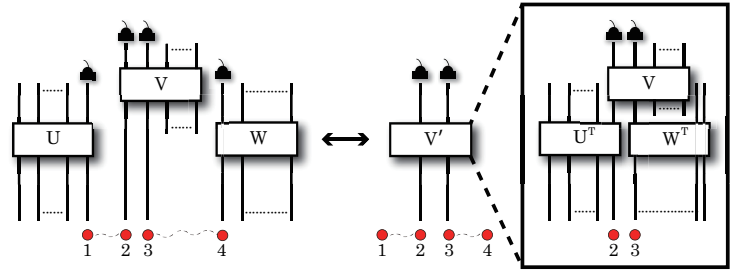}
\end{center}
\captionsetup{justification=raggedright,singlelinecheck=false}
\caption{The equivalence of two optical systems with two Bell pairs as an input. Two red circles connected by a black dashed line represents the Bell pair $|\Phi^+\rangle$. Each of black symbols at output side represents a threshold detector. Each of passive linear optical systems is represented by a corresponding unitary operator $U$, $V$, $W$, $U^T$, or $W^T$. Here, $V'\equiv V(U^T\otimes W^T)$.}
\label{PLO}
\end{figure*}

\subsection{Nonlocal operation for our conversion method}
\label{A}
A general passive linear optical system can be represented as shown in the left-hand side of Fig.~\ref{PLO}, because we consider the situation where nonlocal operations can be performed on only nodes $2$ and $3$, which are close to each other. Note that positions of detectors and input photons can be fixed because swap operations between different spatial modes can be realized using only passive linear optics. In order to simplify the form of the optical system without loss of generality and define our nonlocal operations, we prove the equivalence shown in Fig.~\ref{PLO}. In other words, we show that an output state is the same in both optical systems in Fig.~\ref{PLO} when a postselection succeeds. Here, we assume that the postselection succeeds when all of threshold detectors detect a photon. 

Let $c_{is}^\dag$ and $d_{is}^\dag$ be the $s(\in\{H,V\})$-polarized photon creation operators for input spacial mode $i$ in node $1$ and for output spacial mode $i$ in node $1$, respectively. When a photon in node $1$ is input into the first spacial mode of the optical system shown in the left-hand side of Fig.~\ref{PLO}, $c_{1s}^\dag$ is transformed as follows:
\begin{eqnarray*}
Uc_{1H}^\dag U^\dag&=&\sum_{j=1}^{L'}(u_{jH}d_{jH}^\dag+u_{jV}d_{jV}^\dag)\\
Uc_{1V}^\dag U^\dag&=&\sum_{j=1}^{L'}(\tilde{u}_{jH}d_{jH}^\dag+\tilde{u}_{jV}d_{jV}^\dag),
\end{eqnarray*}
where $L'$ is the number of input (output) special modes. Here, complex numbers $u_{js}$ and $\tilde{u}_{js}$ satisfy $\sum_{j=1}^{L'}(|u_{jH}|^2+|u_{jV}|^2)=\sum_{j=1}^{L'}(|\tilde{u}_{jH}|^2+|\tilde{u}_{jV}|^2)=1$ due to the unitarity of $U$. Accordingly, when the postselection in node $1$ succeeds, i.e., a photon is detected by a threshold detector in node $1$, the Bell pair $|\Phi^+\rangle_{1,2}$ is converted to
\begin{eqnarray}
\nonumber
&&(u_{1H}|H\rangle_1+u_{1V}|V\rangle_1)|H\rangle_2\\
\nonumber
&&+(\tilde{u}_{1H}|H\rangle_1+\tilde{u}_{1V}|V\rangle_1)|V\rangle_2\\
\nonumber
&=&|H\rangle_1(u_{1H}|H\rangle_2+\tilde{u}_{1H}|V\rangle_2)\\
\label{equality}
&&+|V\rangle_1(u_{1V}|H\rangle_2+\tilde{u}_{1V}|V\rangle_2)
\end{eqnarray}
up to normalization. Equation~(\ref{equality}) implies that a state converted from $|\Phi^+\rangle_{1,2}$ in the optical system corresponding to $U_1$ is the same as a state converted from $|\Phi^+\rangle_{1,2}$ in the optical system corresponding to $U_2^T$. For $|\Phi^+\rangle_{3,4}$ and the optical system corresponding to $W$, we can explain in the same way. Furthermore, in order to detect a photon by each of two threshold detectors shown in the right-hand side of Fig.~\ref{PLO}, two photons have to go through the optical systems corresponding to $U_2^T$ and $W_3^T$, respectively. From these facts, we conclude that the equivalence shown in Fig.~\ref{PLO} is satisfied. For our nonlocal conversion methods, it is enough to consider only the optical system for two photons in nodes $2$ and $3$, which corresponds to $V'\equiv V(U^T\otimes W^T)$ (see also the right-hand side of Fig.~\ref{PLO}).

\subsection{Nonlocal four-qubit state conversion}
\label{B}
In this section, we derive success probabilities of optimal nonlocal four-qubit state conversion from $|\Phi^+\rangle_{1,2}|\Phi^+\rangle_{3,4}$ using the optical system given in Sec.~\ref{A}. Without loss of generality, we assume that two photons in nodes $2$ and $3$ are input from spacial modes $1$ and $2$, respectively. After the unitary operator $V'$ is performed on two photons in nodes $2$ and $3$, each of creation operators is transformed as follows:
\begin{eqnarray*}
V'a_{1H}^\dag V'^\dag&=&\sum_{j=1}^L(\beta_{jH}b_{jH}^\dag+\beta_{jV}b_{jV}^\dag)\\
V'a_{1V}^\dag V'^\dag&=&\sum_{j=1}^L(\gamma_{jH}b_{jH}^\dag+\gamma_{jV}b_{jV}^\dag)\\
V'a_{2H}^\dag V'^\dag&=&\sum_{j=1}^L(\alpha_{jH}b_{jH}^\dag+\alpha_{jV}b_{jV}^\dag)\\
V'a_{2V}^\dag V'^\dag&=&\sum_{j=1}^L(\eta_{jH}b_{jH}^\dag+\eta_{jV}b_{jV}^\dag),
\end{eqnarray*}
where $L(\ge 2)$ is the number of input (output) spacial modes. Here, complex numbers $\alpha_{js}$, $\beta_{js}$, $\gamma_{js}$, and $\eta_{js}$ satisfy $\sum_{j=1}^L\sum_{s=H,V}|\alpha_{js}|^2=\sum_{j=1}^L\sum_{s=H,V}|\beta_{js}|^2=\sum_{j=1}^L\sum_{s=H,V}|\gamma_{js}|^2=\sum_{j=1}^L\sum_{s=H,V}|\eta_{js}|^2=1$. When the postselection succeeds, i.e., two photons are detected by two threshold detectors in output spacial modes $1$ and $2$, an output state $|F\rangle$ is given by
\begin{eqnarray*}
|F\rangle&\equiv&\cfrac{\Pi_{\rm post}V'|\Phi^+\rangle_{1,2}|\Phi^+\rangle_{3,4}}{\sqrt{p_{\rm suc}}},
\end{eqnarray*}
where $p_{\rm suc}(>0)$ is the success probability of the postselection, $\Pi_{\rm post}$ is defined as
\begin{eqnarray*}
\Pi_{\rm post}&\equiv&\sum_{s_1=H,V}\sum_{s_2=H,V}\left(\prod_{l=1}^2b_{l{s_l}}^\dag\right)|{\rm vac}\rangle\langle{\rm vac}|\left(\prod_{l'=1}^2b_{l'{s_{l'}}}^\dag\right),\ \ \ \ \ \ \
\end{eqnarray*}
and $|{\rm vac}\rangle$ is a vacuum state. In order to convert to a target state $|t\rangle$, it is necessary to satisfy that $|F\rangle=|t\rangle$. By taking inner product with $|s_1s_2\rangle_{1,4}$, we obtain
\begin{eqnarray}
\label{post_14}
\Pi_{\rm post}V'|s_1s_2\rangle_{2,3}&=&2\sqrt{p_{\rm suc}}\langle s_1s_2|_{1,4}|t\rangle.
\end{eqnarray}
By substituting $(s_1,s_2)=(H,H)$,  $(H,V)$, $(V,H)$, and $(V,V)$ into Eq.~(\ref{post_14}), we obtain following four equations:
\begin{eqnarray}
\label{C1}
\Pi_{\rm post}V'|HH\rangle_{2,3}&=&2\sqrt{p_{\rm suc}}\langle HH|_{1,4}|t\rangle_{1,2,3,4}\\
\label{C2}
\Pi_{\rm post}V'|HV\rangle_{2,3}&=&2\sqrt{p_{\rm suc}}\langle HV|_{1,4}|t\rangle_{1,2,3,4}\\
\label{C3}
\Pi_{\rm post}V'|VH\rangle_{2,3}&=&2\sqrt{p_{\rm suc}}\langle VH|_{1,4}|t\rangle_{1,2,3,4}\\
\label{C4}
\Pi_{\rm post}V'|VV\rangle_{2,3}&=&2\sqrt{p_{\rm suc}}\langle VV|_{1,4}|t\rangle_{1,2,3,4}.
\end{eqnarray}
Furthermore, by taking inner product with $|s_1s_2\rangle_{2,3}$ in both sides of each of Eqs.~(\ref{C1}), (\ref{C2}), (\ref{C3}), and (\ref{C4}), we obtain 16 constraints in total. Considering these constraints, we derive the optimal value of the success probability $p_{\rm suc}$ for well-known four-qubit entangled states (See Appendix {\red A} for details). The optimal success probabilities are summarized in Table~\ref{T1}.

\begin{table}[t]
\begin{tabular}{|c|c|} \hline
    $|t\rangle$ & Optimal value of $p_{\rm suc}$ \\ \hline
    $|C_4\rangle$ & $1/9$ \\ \hline
    $|GHZ_4\rangle$ & $1/2$ \\ \hline
    $|W_4\rangle$ & $0$ \\ \hline
    $|D_4^{(2)}\rangle$ & $0$ \\ \hline
    $|\Phi^+\rangle_{1,3}|\Phi^+\rangle_{2,4}$ & $1$ \\ \hline
    $|\Phi^+\rangle_{1,4}|\Phi^+\rangle_{2,3}$ & $1/4$ \\ \hline
    $|\chi\rangle$ & $0$ \\ \hline
\end{tabular}
\captionsetup{justification=raggedright,singlelinecheck=false}
\caption{Optimal values of $p_{\rm suc}$ for well-known four-qubit entangled states.}
\label{T1}
\end{table}
\begin{figure*}[t]
\begin{center}
\includegraphics[width=12cm, clip]{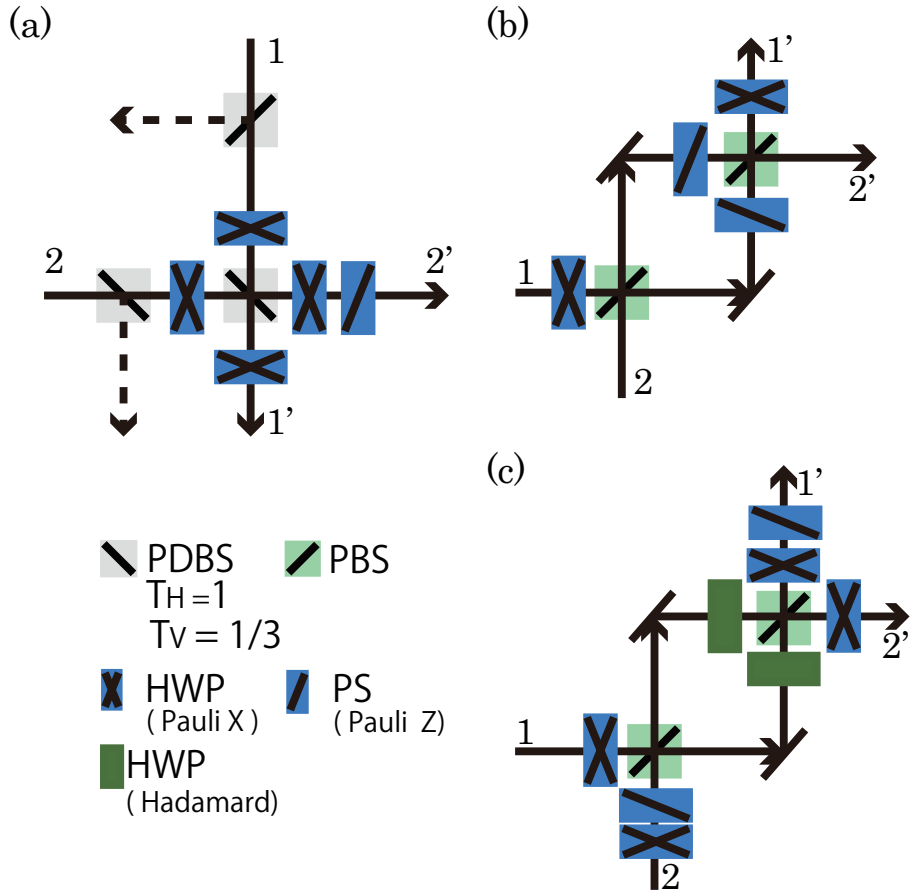}
\end{center}
\captionsetup{justification=raggedright,singlelinecheck=false}
\caption{Optical systems that achieve optimal values of $p_{\rm suc}$. If one photon is detected in each of output special modes $1$ and $2$, which are denoted by $1'$ and $2'$, respectively, the postselection succeeds. ({\bf a}) A nonlocal converter for $|C_4\rangle$. {\red If one photon goes for either of two dashed arrows, the conversion is failed.} ({\bf b}) A nonlocal converter for $|GHZ_4\rangle$. ({\bf c}) A nonlocal converter for $|\Phi^+\rangle_{1,4}|\Phi^+\rangle_{2,3}$. Note that photons proceed in the direction of the arrows. A gray box with a diagonal line represents PDBS whose transmittance for H(V)-polarized photons is $1$ $(1/3)$. A blue rectangle with two diagonal lines represents a wave plate that operates as $X$. A blue rectangle with a diagonal line represents a PS that operates as $Z$. A black line represents a mirror. A pale green box with a diagonal line represents PDBS whose transmittance for H(V)-polarized photons is $1$ $(0)$, i.e., polarizing beam splitter (PBS). A green rectangle represents a wave plate that operates as the Hadamard gate.}
\label{optics}
\end{figure*}

Next, we consider optical systems that achieve the optimal values of $p_{\rm suc}$ (See Fig.~\ref{optics}). In fact, the optical systems given in Fig.~\ref{optics} are equivalent to existing optical systems proposed for well-known four-qubit entangled states, which are given in Table~\ref{T1}. The optical system shown in Fig.~\ref{optics} ({\bf a}) is essentially equivalent to the controlled-$Z$ gate and the controlled-$X$ gate proposed in Refs.~\cite{[KSWUW05],[OHTS05]}. The optical systems shown in Fig.~\ref{optics} ({\bf b}) and ({\bf c}) are also essentially equivalent to a nonlocal gate proposed in Ref.~\cite{[TTONKW16]}. With respect to an optical system for conversion to $|\Phi^+\rangle_{1,3}|\Phi^+\rangle_{2,4}$, which is not shown here, it can obviously be constructed using only mirrors. Note that these optical systems are not unique ones that achieve optimal success probabilities. In fact, $|GHZ_4\rangle$ can also be converted from $|\Phi^+\rangle_{1,2}|\Phi^+\rangle_{3,4}$ with probability $1/2$ by performing the type-II fusion gate, which is not shown here~\cite{[PDGWZ01]}. While it was not known whether these existing optical systems are optimal or not, our results show the optimality of these existing optical systems from a network perspective. Our results also show that $|W_4\rangle$, $|D_4^{(2)}\rangle$, and $|\chi\rangle$ cannot be converted from $|\Phi^+\rangle_{1,2}|\Phi^+\rangle_{3,4}$ using only passive linear optics and postselection (see Table~\ref{T1}).

In the last of this section, we analyze practicality of our schemes. First, we focus on the optical system shown in Fig.~\ref{optics} ({\bf a}). We consider how change in transmittance of PDBSs affects the success probability $p_{\rm suc}$ and the fidelity $F$ between an output quantum state and $|C_4\rangle$. For simplicity, we assume that all of PDBSs have the same transmittance $T_H$ and $T_V$ for $H$- and $V$-polarized {\red photons}, respectively. In the ideal case, $T_H$ and $T_V$ is set to $1$ and $1/3$, respectively. Furthermore, we assume that HWPs and the PS ideally work because we can adjust their function by rotating angles of them. The efficiency of two threshold detectors are denoted by $\eta$ and $\eta'$. In this case, $p_{\rm suc}$ and $F$ are written as
{\red\begin{eqnarray}
\nonumber
&&p_{\rm suc}\\
\label{p_suc_im}
&=&\cfrac{\eta\eta'(T_H^2+2T_HT_V+T_V^2-6T_H^2T_V-6T_HT_V^2+12T_H^2T_V^2)}{4}\ \ \ \ \ \ \ \ \\
\nonumber
&&F\\
\label{fidelity_im}
&=&\cfrac{\left|T_H+2T_HT_V-T_V\right|}{2\sqrt{T_H^2+2T_HT_V+T_V^2-6T_H^2T_V-6T_HT_V^2+12T_H^2T_V^2}}
\end{eqnarray}}
and are plotted in Fig.~\ref{error} in the case of $\eta=\eta'=1$ {\red (See Appendix B for the derivation of $p_{\rm suc}$ and $F$)}. Accordingly, the efficiency of detectors does not affect the fidelity. Furthermore, from Fig.~\ref{error} ({\bf b}), it is known that $F\ge 0.9$ is satisfied even when the deviation from the ideal values of $T_H$ and $T_V$ is {\red$0.14$}, i.e., {\red$T_H=0.86$} and {\red$T_V=(1\pm0.14)/3$}. This implies that our scheme is robust against experimental imperfections and detector inefficiencies. With respect to photon distinguishability, we can adopt the same analysis used in Ref.~\cite{[KSWUW05]}. As a result, considerable deviation from the optimal performance occurs due to photon distinguishability as with Ref.~\cite{[KSWUW05]}.
Second, we focus on the optical systems shown in Fig.~\ref{optics} ({\bf b}) and ({\bf c}). Since these optical systems are based on the Mach-Zehnder interferometer, stabilization is important to realize them. By using a Sagnac interferometer, we can improve stability of them. In fact, essentially equivalent optical systems have already been experimentally characterized by using the Sagnac interferometer~\cite{[MSMMSJTOT17]}. Accordingly, by using the same technique, it would be possible to realize our method.

\begin{figure*}[t]
\begin{center}
\includegraphics[width=12cm, clip]{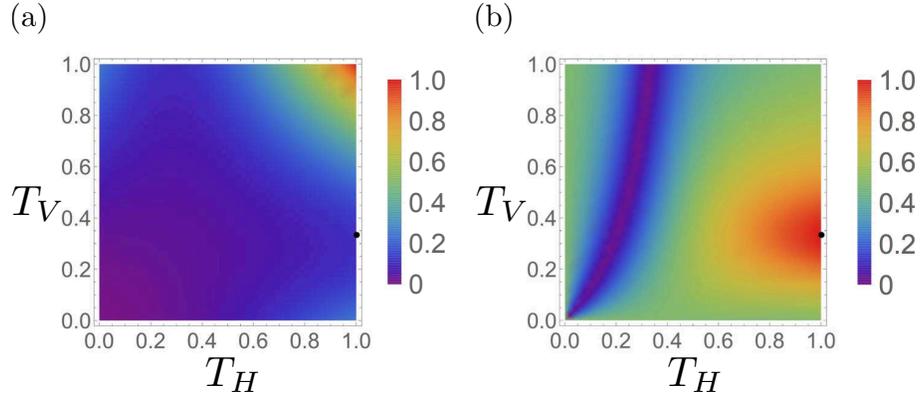}
\end{center}
\captionsetup{justification=raggedright,singlelinecheck=false}
\caption{The transmittance-dependence of the success probability and the fidelity of the optical system shown in Fig.~\ref{optics} ({\bf a}). {\red $T_H$ and $T_V$ denote the transmittance of PDBSs for $H$- and $V$-polarized photons, respectively.} The black dot represents the ideal case. ({\bf a}) The success probability. ({\bf b}) The fidelity between an output quantum state and $|C_4\rangle$.}
\label{error}
\end{figure*}

\subsection{Improvement of success probabilities using ancillary qubits}
\label{C}
\begin{figure}[t]
\begin{center}
\includegraphics[width=8cm, clip]{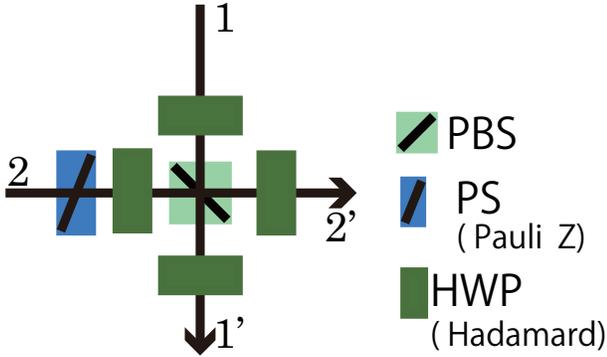}
\end{center}
\captionsetup{justification=raggedright,singlelinecheck=false}
\caption{An optical system that converts to $|\chi\rangle$ from $|C_4\rangle$ if one photon is detected in each of output spacial modes $1$ and $2$, which are denoted by $1'$ and $2'$, respectively.}
\label{chi}
\end{figure}

\begin{figure*}[t]
\begin{center}
\includegraphics[width=12cm, clip]{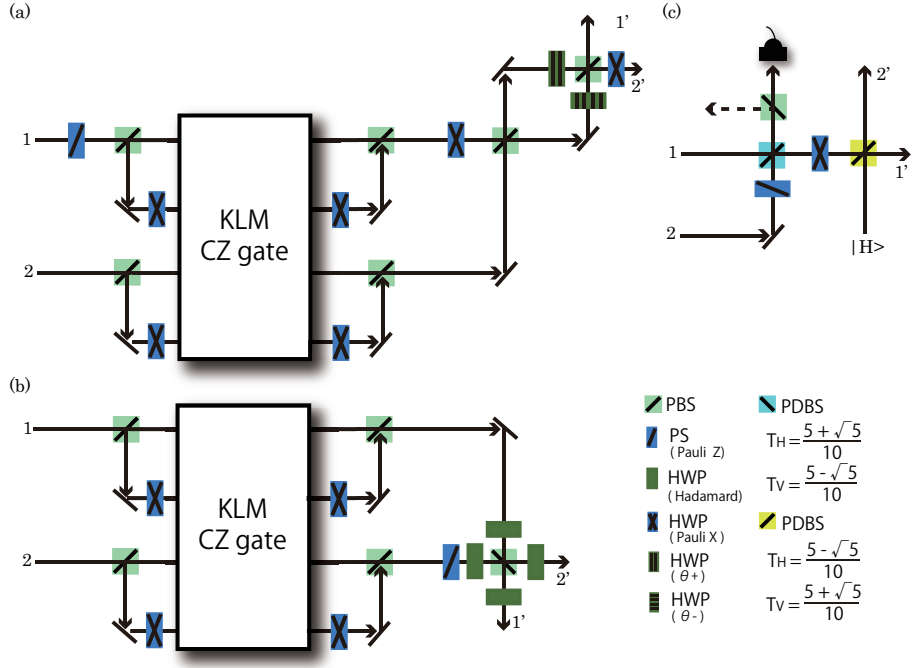}
\end{center}
\captionsetup{justification=raggedright,singlelinecheck=false}
\caption{Optical systems considered in Sec.~\ref{C}. The KLM CZ gate represents a optical system given in Fig. 2 of Ref.~\cite{[KLM01]}. The HWP ($\theta_\pm$) converts $|H\rangle$ $(|V\rangle)$ into $\cos{(2\theta_\pm)}|H\rangle+\sin{(2\theta_\pm)}|V\rangle$ $(\sin{(2\theta_\pm)}|H\rangle-\cos{(2\theta_\pm)}|V\rangle)$, where $2\theta_\pm\equiv\arcsin{\sqrt{(5\pm\sqrt{5})/10}}$. ({\bf a}) A nonlocal converter for $|D_4^{(2)}\rangle$. ({\bf b}) A nonlocal converter for $|\chi\rangle$. ({\bf c}) A nonlocal converter for $|W_4\rangle$. When one photon is detected by a threshold detector, this converter correctly works. {\red On the other hand, if at least one photon goes for the dashed arrow, the conversion is failed.}}
\label{gate}
\end{figure*}

We show that $|W_4\rangle$, $|D_4^{(2)}\rangle$, and $|\chi\rangle$ can be probabilistically converted from $|\Phi^+\rangle_{1,2}|\Phi^+\rangle_{3,4}$ if ancillary qubits are available in addition to passive linear optics and postselection. Note that we minimize the access to ancillary qubits as possible. To this end, we also consider the conversion from $|C_4\rangle$ to $|D_4^{(2)}\rangle$ and $|\chi\rangle$. Such conversion can be done with probabilities $3/10$ and $1/2$ using optical systems given in Ref.~\cite{[TTONKW16]} and Fig.~\ref{chi}, respectively.
Based on this fact, we construct conversion methods for $|D_4^{(2)}\rangle$ and $|\chi\rangle$ as follows: First, polarization encoded qubits are converted into spatial dual-rail qubits using a PDBS and a wave plate. This can be done with probability $1$. Second, the controlled-$Z$ gate is performed on photons in nodes $2$ and $3$ using Knill-Laflamme-Milburn (KLM) scheme~\cite{[KLM01]}. This step requires two ancillary qubits and succeeds with probability $1/16$. Third, spatial dual-rail qubits are converted into polarization-encoded qubits using a PDBS and a wave plate. As a result, $|\Phi^+\rangle_{1,2}|\Phi^+\rangle_{3,4}$ is converted to $|C_4\rangle$. Finally, $|C_4\rangle$ is converted to $|D_4^{(2)}\rangle$ or $|\chi\rangle$ using the optical systems mentioned above. Accordingly, $|\Phi^+\rangle_{1,2}|\Phi^+\rangle_{3,4}$ can be converted to $|D_4^{(2)}\rangle$ and $|\chi\rangle$ using two ancillary qubits with probabilities $3/160$ and $1/32$, respectively. The optical systems for $|D_4^{(2)}\rangle$ and $|\chi\rangle$ are given in Fig.~\ref{gate} ({\bf a}) and ({\bf b}), respectively.
For $|W_4\rangle$, we can also construct  a conversion method based on existing methods as follows:
First, we transform $|\Phi^+\rangle_{1,2}|\Phi^+\rangle_{3,4}$ to $|W_3\rangle$ using method proposed in Ref.~\cite{[TWOYKI09]}. This transformation requires no ancillary qubit and succeeds with probability $3/20$. Next we use the expansion method proposed in Refs.~\cite{[TOYKI09],[ITYKI11]} to generate $|W_4\rangle$ from $|W_3\rangle$ and one ancillary qubit. This expansion succeeds with probability $4/15$. Accordingly, this conversion method succeeds with $1/25$ and requires only one ancillary qubit. The optical system for $|W_4\rangle$ is given in Fig.~\ref{gate} ({\bf c}).


\section{Comparison with previous schemes}
\label{add}
In previous conversion schemes using only LOCC, $|\Phi^+\rangle_{1,2}|\Phi^+\rangle_{3,4}$ cannot be converted to $|C_4\rangle$, $|GHZ_4\rangle$, $|\Phi^+\rangle_{1,3}|\Phi^+\rangle_{2,4}$, and $|\Phi^+\rangle_{1,4}|\Phi^+\rangle_{2,3}$. In our optimal scheme, we realize such conversion using nonlocal two-qubit operations without ancillary qubits. Furthermore, in the situation considered in Secs.~\ref{II} and \ref{B}, our scheme is optimal.

We also compare our optimal scheme given in Sec.~\ref{B} with a conversion scheme based on the KLM scheme~\cite{[KLM01]}, which is a scheme for universal quantum computing, in terms of the number of required ancillary qubits and the success probability. Note that we consider the KLM scheme that succeeds with probability $1/16$ and requires two ancillary qubits to perform the controlled-$Z$ gate. In order to convert to $|C_4\rangle$ from $|\Phi^+\rangle_{1,2}|\Phi^+\rangle_{3,4}$, one controlled-$Z$ gate is required. Accordingly, the KLM-based scheme requires two ancillary qubits and succeeds with probability $1/16$. In order to convert to $|GHZ_4\rangle$, $|\Phi^+\rangle_{1,3}|\Phi^+\rangle_{2,4}$, or $|\Phi^+\rangle_{1,4}|\Phi^+\rangle_{2,3}$ from $|\Phi^+\rangle_{1,2}|\Phi^+\rangle_{3,4}$, three controlled-$Z$ gates are required. As a result, the KLM-based scheme requires six ancillary qubits and succeeds with probability $1/4096$. This argument implies that our method is more efficient than the KLM-based scheme in terms of the number of required ancillary qubits and the success probability when $|\Phi^+\rangle_{1,2}|\Phi^+\rangle_{3,4}$ is converted to $|C_4\rangle$, $|GHZ_4\rangle$, $|\Phi^+\rangle_{1,3}|\Phi^+\rangle_{2,4}$, or $|\Phi^+\rangle_{1,4}|\Phi^+\rangle_{2,3}$. On the other hand, when $|\Phi^+\rangle_{1,2}|\Phi^+\rangle_{3,4}$ is converted to $|W_4\rangle$, $|D_4^{(2)}\rangle$, or $|\chi\rangle$, the KLM-based scheme is more efficient than our proposed scheme because such conversion cannot be achieved using our scheme.

\section{Conclusion}
\label{IV}
In this work, we have considered two kinds of two-qubit operations to analyze optimal nonlocal conversion of two Bell pairs. First, we have derived a necessary and sufficient condition for two-qubit unitary operators. Second, we have derived optimal success probabilities for nonlocal conversion using passive linear optical systems and a postselection. In these arguments, we have assumed that no ancillary qubits are available. Furthermore, we have shown that the optimal success probabilities can be improved using ancillary qubits. Finally, we compare our scheme with previously proposed schemes. Our method is more efficient than the KLM-based scheme when $|\Phi^+\rangle_{1,2}|\Phi^+\rangle_{3,4}$ is converted to $|C_4\rangle$, $|GHZ_4\rangle$, $|\Phi^+\rangle_{1,3}|\Phi^+\rangle_{2,4}$, or $|\Phi^+\rangle_{1,4}|\Phi^+\rangle_{2,3}$. On the other hand, when $|\Phi^+\rangle_{1,2}|\Phi^+\rangle_{3,4}$ is converted to $|W_4\rangle$, $|D_4^{(2)}\rangle$, or $|\chi\rangle$, the KLM-based scheme is more efficient than our scheme.

In the quantum networks, we have various situations for sharing entangled resource states. For example, if nonlocal operations performed on three photons are available, $|C_4\rangle$ can be converted from $|\Phi^+\rangle|HH\rangle$ with probability $1/4$~\cite{[TKYKI08]}. This success probability is greater than one that can be achieved by our method. As another example, an optical system proposed in Ref.~\cite{[WRRSWVAZ05]} can also be used for state conversion from superposition of two Bell pairs to several four-qubit graph states. It is an important step to clarify what we can do using nonlocal operations towards the realization of advanced quantum networks. Thus, there still remains an interesting open problem of how to optimally convert to various entangled resource states with a certain initial state and restricted quantum operations in quantum networks.

\medskip
\begin{center}
{\bf ACKNOWLEDGMENTS}
\end{center}

Y.T. is supported by the Program for Leading Graduate Schools: Interactive Materials Science Cadet Program and JSPS Grant-in-Aid for JSPS Research fellow No.JP17J03503.
N.I. is supported by CREST, JST JPMJCR1671 and JSPS Grant-in-Aid for Scientific Research(A) JP16H02214.

\medskip
\section*{\red APPENDIX A}
We derive the achievable maximum values of $p_{\rm suc}$ for the conversion from the initial state $|\Phi^+\rangle_{1,2}|\Phi^+\rangle_{3,4}$ to the  well-known four-qubit entangled states, such as $|C_4\rangle$, $|{GHZ}_4\rangle$, $|W_4\rangle$, $|D_4^{(2)}\rangle$, $|\Phi^+\rangle_{1,3}|\Phi^+\rangle_{2,4}$, $|\Phi^+\rangle_{1,4}|\Phi^+\rangle_{2,3}$, and $|\chi\rangle$.

\setcounter{subsection}{0}

\subsection{$|t\rangle=|C_4\rangle$}
The success probability of the state conversion for the state $|C_4\rangle$ is calculated with 
\begin{eqnarray*}
\Pi_{\rm post}V'|\Phi^+\rangle_{1,2}|\Phi^+\rangle_{3,4}=\sqrt{p_{\rm suc}}|C_4\rangle.
\end{eqnarray*}
From inner products of $|s_1s_2\rangle_{1,4}$ $(s_1,s_2\in\{H,V\})$, we obtain
\begin{eqnarray}
\label{179}
\Pi_{\rm post}V'|HH\rangle_{2,3}&=&\sqrt{p_{\rm suc}}|HH\rangle_{2,3},\\
\label{180}
\Pi_{\rm post}V'|HV\rangle_{2,3}&=&\sqrt{p_{\rm suc}}|HV\rangle_{2,3},\\
\label{181}
\Pi_{\rm post}V'|VH\rangle_{2,3}&=&\sqrt{p_{\rm suc}}|VH\rangle_{2,3},
\end{eqnarray}
and
\begin{eqnarray}
\label{182}
\Pi_{\rm post}V'|VV\rangle_{2,3}&=&-\sqrt{p_{\rm suc}}|VV\rangle_{2,3}.
\end{eqnarray}
Using inner products of $|s_1s_2\rangle_{2,3}$ in both sides of each of Eqs.~(\ref{179}), (\ref{180}), (\ref{181}), and (\ref{182}), we also obtain
\begin{eqnarray}
\label{183}
\sqrt{p_{\rm suc}}&=&\beta_{1H}\alpha_{2H}+\beta_{2H}\alpha_{1H}\\
\label{184}
&=&\beta_{1H}\eta_{2V}+\beta_{2V}\eta_{1H}\\
\label{185}
&=&\gamma_{1V}\alpha_{2H}+\gamma_{2H}\alpha_{1V}\\
\label{186}
&=&-(\gamma_{1V}\eta_{2V}+\gamma_{2V}\eta_{1V}),
\end{eqnarray}
and
\begin{eqnarray}
\label{187}
0&=&\beta_{1H}\alpha_{2V}+\beta_{2V}\alpha_{1H}\\
\label{188}
&=&\beta_{1V}\alpha_{2H}+\beta_{2H}\alpha_{1V}\\
\label{189}
&=&\beta_{1V}\alpha_{2V}+\beta_{2V}\alpha_{1V}\\
\nonumber
&=&\beta_{1H}\eta_{2H}+\beta_{2H}\eta_{1H}\\
\nonumber
&=&\beta_{1V}\eta_{2H}+\beta_{2H}\eta_{1V}\\
\nonumber
&=&\beta_{1V}\eta_{2V}+\beta_{2V}\eta_{1V}\\
\nonumber
&=&\gamma_{1H}\alpha_{2H}+\gamma_{2H}\alpha_{1H}\\
\nonumber
&=&\gamma_{1H}\alpha_{2V}+\gamma_{2V}\alpha_{1H}\\
\label{195}
&=&\gamma_{1V}\alpha_{2V}+\gamma_{2V}\alpha_{1V}\\
\label{196}
&=&\gamma_{1H}\eta_{2H}+\gamma_{2H}\eta_{1H}\\
\label{197}
&=&\gamma_{1H}\eta_{2V}+\gamma_{2V}\eta_{1H}\\
\label{198}
&=&\gamma_{1V}\eta_{2H}+\gamma_{2H}\eta_{1V}.
\end{eqnarray}
From Eqs.~(\ref{183}), (\ref{187}), (\ref{188}), and (\ref{189}),
\begin{widetext}
\begin{eqnarray}
\label{22}
0&=&(\beta_{1V}\alpha_{2H}+\beta_{2H}\alpha_{1V})\alpha_{1H}\alpha_{2V}+(\beta_{1H}\alpha_{2V}+\beta_{2V}\alpha_{1H})\alpha_{2H}\alpha_{1V}\\
\nonumber
&=&(\beta_{1V}\alpha_{2V}+\beta_{2V}\alpha_{2H})\alpha_{1H}\alpha_{2V}+(\beta_{1H}\alpha_{2H}+\beta_{2H}\alpha_{1H})\alpha_{1V}\alpha_{2V}\\
\nonumber
&=&\sqrt{p_{\rm suc}}\alpha_{1V}\alpha_{2V}.
\end{eqnarray}
\end{widetext}
Here, $\alpha_{1V}\alpha_{2V}=0$ is required to satisfy $p_{\rm suc}>0$. Replacing  $\alpha$ with $\beta$ in Eq.~(\ref{22}), $\beta_{1V}\beta_{2V}=0$ is also required. Moreover, from the same way with Eqs.~(\ref{186}), (\ref{196}), (\ref{197}), and (\ref{198}),  $\eta_{1H}\eta_{2H}=\gamma_{1H}\gamma_{2H}=0$. If $\alpha_{2V}\neq0$, $\alpha_{1V}=\gamma_{1V}=0$ is derived from $\alpha_{1V}\alpha_{2V}=0$ and Eq.~(\ref{195}), and then $p_{\rm suc}=0$ is derived from Eq.~(\ref{185}). In order to satisfy $p_{\rm suc}>0$, $\alpha_{2V}=0$ has to be satisfied. According to the same way used in the above proof, it also requires $\beta_{1V}=\eta_{2H}=\gamma_{1H}=0$. By substituting $\alpha_{2V}=0$ into Eq.~(\ref{187}), we obtain $0=\beta_{2V}\alpha_{1H}$. Here, we consider three cases: (I) $\beta_{2V}=0$ and $\alpha_{1H}\neq0$, (II) $\beta_{2V}\neq0$ and $\alpha_{1H}=0$, and (III) $\beta_{2V}=\alpha_{1H}=0$. 

First, we consider the case (I). From Eqs.~(\ref{183})-(\ref{186}), the square root of the success probability is given by
\begin{eqnarray*}
\sqrt{p_{\rm suc}}&=&\beta_{1H}\alpha_{2H}+\beta_{2H}\alpha_{1H}\\
&=&\beta_{1H}\eta_{2V}\\
&=&\gamma_{1V}\alpha_{2H}\\
&=&-\gamma_{1V}\eta_{2V}.
\end{eqnarray*}
As a result, $\alpha_{2H}=-\eta_{2V}$ and $\beta_{1H}=-\gamma_{1V}$ are satisfied, and they imply that $\beta_{2H}\alpha_{1H}=-2\beta_{1H}\alpha_{2H}$. Since $|\beta_{1H}|^2+|\beta_{2H}|^2\le1$ and $|\alpha_{1H}|^2+|\alpha_{2H}|^2\le1$ are satisfied from $\sum_{j=1}^L\sum_{s=H,V}|{\alpha}_{js}|^2=\sum_{j=1}^L\sum_{s=H,V}|\beta_{js}|^2=1$, the relation is rewritten as follows: 
\begin{eqnarray}
\nonumber
4|\beta_{1H}\alpha_{2H}|^2&=&|\beta_{2H}\alpha_{1H}|^2\\
\nonumber
&\le&(1-|\beta_{1H}|^2)(1-|\alpha_{2H}|^2)\\
\label{217}
|\beta_{1H}\alpha_{2H}|^2&\le&\cfrac{1-(|\alpha_{2H}|^2+|\beta_{1H}|^2)}{3}.
\end{eqnarray}
Since $p_{\rm suc}=|\beta_{1H}\alpha_{2H}|^2$ is a monotone increasing function of $|\beta_{1H}|$ and $|\alpha_{2H}|$, and $[1-(|\alpha_{2H}|^2+|\beta_{1H}|^2)]/3$ is a monotone decreasing function of $|\beta_{1H}|$ and $|\alpha_{2H}|$, $p_{\rm suc}$ is maximized by $|\beta_{1H}|$ and $|\alpha_{2H}|$ that satisfy equality in Eq.~(\ref{217}). We define $\epsilon(>0)$ such that $|\beta_{1H}|^2=\epsilon|\alpha_{2H}|^2$ holds. From above calculation, an upper bound of the success probability is given by
\begin{eqnarray}
\label{32}
p_{\rm suc}\le\underset{\epsilon}{\rm max}\left\{\cfrac{2\epsilon^2+16\epsilon+2-2(1+\epsilon)\sqrt{\epsilon^2+14\epsilon+1}}{36\epsilon}\right\}.\ \ \
\end{eqnarray}
Since the right-hand side of Eq.~(\ref{32}) is maximized with $\epsilon=1$, $p_{\rm suc}\le1/9$. For the case (II), according to the process in the same way as well as (I), $p_{\rm suc}\le1/9$. When we consider the case (III), it can be divided into four cases: (i) $\beta_{2H}=\gamma_{2H}=\gamma_{2V}=0$, (ii) $\alpha_{1V}=\eta_{1H}=\eta_{1V}=0$, (iii) $\beta_{2H}=\eta_{1H}=\eta_{1V}=0$, and (iv) $\beta_{2H}=\alpha_{1V}=\eta_{1H}=\gamma_{2H}=0$. For (i) and (ii), from Eqs.~(\ref{183})--(\ref{186}), $\eta_{2V}=\alpha_{2H}=-\eta_{2V}$ is satisfied. As a result, $p_{\rm suc}=0$. For (iii) and (iv), with the same way as well as (I), $p_{\rm suc}\le1/9$.

\subsection{$|t\rangle=|GHZ_4\rangle$}
The conversion from  $|\Phi^+\rangle_{1,2}|\Phi^+\rangle_{3,4}$ to the state $|GHZ_4\rangle$ is written by
\begin{eqnarray*}
\Pi_{\rm post}V'|\Phi^+\rangle_{1,2}|\Phi^+\rangle_{3,4}=\sqrt{p_{\rm suc}}|GHZ_4\rangle.
\end{eqnarray*}
We also obtain the following equations from inner products of $|s_1s_2\rangle_{1,4}$:
\begin{eqnarray}
\label{34}
\Pi_{\rm post}V'|HH\rangle_{2,3}&=&\sqrt{2p_{\rm suc}}|HH\rangle_{2,3},\\
\label{35}
\Pi_{\rm post}V'|HV\rangle_{2,3}&=&0,\\
\label{36}
\Pi_{\rm post}V'|VH\rangle_{2,3}&=&0,
\end{eqnarray}
and
\begin{eqnarray}
\label{37}
\Pi_{\rm post}V'|VV\rangle_{2,3}&=&\sqrt{2p_{\rm suc}}|VV\rangle_{2,3}.
\end{eqnarray}
We give the following relations from inner products of $|s_1s_2\rangle_{2,3}$ with Eqs.~(\ref{34}), (\ref{35}), (\ref{36}), and (\ref{37}): 
\begin{eqnarray}
\label{140}
\sqrt{2p_{\rm suc}}&=&\beta_{1H}\alpha_{2H}+\beta_{2H}\alpha_{1H}\\
\label{141}
&=&\gamma_{1V}\eta_{2V}+\gamma_{2V}\eta_{1V},
\end{eqnarray}
and
\begin{eqnarray}
\label{142}
0&=&\beta_{1H}\alpha_{2V}+\beta_{2V}\alpha_{1H}\\
\label{143}
&=&\beta_{1V}\alpha_{2H}+\beta_{2H}\alpha_{1V}\\
\label{144}
&=&\beta_{1V}\alpha_{2V}+\beta_{2V}\alpha_{1V}\\
\label{145}
&=&\beta_{1H}\eta_{2H}+\beta_{2H}\eta_{1H}\\
\label{146}
&=&\beta_{1H}\eta_{2V}+\beta_{2V}\eta_{1H}\\
\nonumber
&=&\beta_{1V}\eta_{2H}+\beta_{2H}\eta_{1V}\\
\nonumber
&=&\beta_{1V}\eta_{2V}+\beta_{2V}\eta_{1V}\\
\nonumber
&=&\gamma_{1H}\alpha_{2H}+\gamma_{2H}\alpha_{1H}\\
\nonumber
&=&\gamma_{1H}\alpha_{2V}+\gamma_{2V}\alpha_{1H}\\
\nonumber
&=&\gamma_{1V}\alpha_{2H}+\gamma_{2H}\alpha_{1V}\\
\nonumber
&=&\gamma_{1V}\alpha_{2V}+\gamma_{2V}\alpha_{1V}\\
\label{153}
&=&\gamma_{1H}\eta_{2H}+\gamma_{2H}\eta_{1H}\\
\label{154}
&=&\gamma_{1H}\eta_{2V}+\gamma_{2V}\eta_{1H}\\
\label{155}
&=&\gamma_{1V}\eta_{2H}+\gamma_{2H}\eta_{1V}.
\end{eqnarray}
From Eqs.~(\ref{140}), (\ref{142}), (\ref{143}), and (\ref{144}),
\begin{widetext}
\begin{eqnarray}
\label{54}
0&=&(\beta_{1V}\alpha_{2H}+\beta_{2H}\alpha_{1V})\alpha_{1H}\alpha_{2V}+(\beta_{1H}\alpha_{2V}+\beta_{2V}\alpha_{1H})\alpha_{2H}\alpha_{1V}\\
\nonumber
&=&\sqrt{2p_{\rm suc}}\alpha_{1V}\alpha_{2V}.
\end{eqnarray}
\end{widetext}
Accordingly,  $\alpha_{1V}\alpha_{2V}=0$ to satisfy $p_{\rm suc}>0$. By replacing $\alpha$ with $\beta$ in Eq.~(\ref{54}), we also obtain $\beta_{1V}\beta_{2V}=0$. 
As results of performing the similar calculation for Eqs.~(\ref{141}), (\ref{153}), (\ref{154}), and (\ref{155}), we get the relation of $\eta_{1H}\eta_{2H}=\gamma_{1H}\gamma_{2H}=0$. From above results,  $\alpha_{1V}=0$ or $\alpha_{2V}=0$ must be satisfied at least. For $\alpha_{1V}$ and $\alpha_{2V}$, here, we redefine $\alpha_{iV}$ as $\alpha_{iV}=0$ and $\alpha_{\overline{i}V}$. For $\eta_{1H}$ and $\eta_{2H}$, we also redefine $\eta_{kH}$ as $\eta_{kH}=0$ and $\eta_{\overline{k}H}$. From Eqs.~(\ref{145}) and (\ref{146}) and above redefinitions,
\begin{eqnarray}
\label{156}
0=\beta_{kH}\eta_{\overline{k}H}=\beta_{kH}\eta_{\overline{k}V}.
\end{eqnarray}
Accordingly, we then have to consider two cases: (I) $\beta_{kH}=0$ and (II) $\beta_{kH}\neq0$.
\begin{enumerate}
\item[(I)] $\beta_{kH}=0$\\
From Eq.~(\ref{140}),
\begin{eqnarray*}
p_{\rm suc}=\cfrac{|\beta_{\overline{k}H}|^2|\alpha_{kH}|^2}{2}\le\cfrac{1}{2}.
\end{eqnarray*}
\item[(II)] $\beta_{kH}\neq0$\\
From Eq.~(\ref{156}), $\eta_{\overline{k}H}=\eta_{\overline{k}V}=0$ is derived and then
\begin{eqnarray*}
p_{\rm suc}=\cfrac{|\gamma_{\overline{k}V}|^2|\eta_{kV}|^2}{2}\le\cfrac{1}{2}.
\end{eqnarray*}
\end{enumerate}
Therefore, the maximum value of the success probability is given by $p_{\rm suc}\le 1/2$.

\subsection{$|t\rangle=|W_4\rangle$}
The conversion from  $|\Phi^+\rangle_{1,2}|\Phi^+\rangle_{3,4}$ to the state $|W_4\rangle$ is given by
\begin{eqnarray*}
\Pi_{\rm post}V'|\Phi^+\rangle_{1,2}|\Phi^+\rangle_{3,4}=\sqrt{p_{\rm suc}}|W_4\rangle.
\end{eqnarray*}
By taking inner products of $|s_1s_2\rangle_{1,4}$, we obtain
\begin{eqnarray}
\label{60}
\Pi_{\rm post}V'|HH\rangle_{2,3}&=&\sqrt{p_{\rm suc}}(|HV\rangle_{2,3}+|VH\rangle_{2,3}),\\
\label{61}
\Pi_{\rm post}V'|HV\rangle_{2,3}&=&\sqrt{p_{\rm suc}}|HH\rangle_{2,3},\\
\label{62}
\Pi_{\rm post}V'|VH\rangle_{2,3}&=&\sqrt{p_{\rm suc}}|HH\rangle_{2,3},
\end{eqnarray}
and
\begin{eqnarray}
\label{63}
\Pi_{\rm post}V'|VV\rangle_{2,3}=0.
\end{eqnarray}
From inner products of $|s_1s_2\rangle_{2,3}$ with Eqs.~(\ref{60}), (\ref{61}), (\ref{62}), and (\ref{63}), we obtain
\begin{eqnarray}
\label{119}
\sqrt{p_{\rm suc}}&=&\beta_{1H}\alpha_{2V}+\beta_{2V}\alpha_{1H}\\
\label{120}
&=&\beta_{1V}\alpha_{2H}+\beta_{2H}\alpha_{1V}\\
\label{121}
&=&\beta_{1H}\eta_{2H}+\beta_{2H}\eta_{1H}\\
\label{122}
&=&\gamma_{1H}\alpha_{2H}+\gamma_{2H}\alpha_{1H},
\end{eqnarray}
and 
\begin{eqnarray}
\nonumber
0&=&\beta_{1H}\alpha_{2H}+\beta_{2H}\alpha_{1H}\\
\label{124}
&=&\beta_{1V}\alpha_{2V}+\beta_{2V}\alpha_{1V}\\
\label{125}
&=&\beta_{1H}\eta_{2V}+\beta_{2V}\eta_{1H}\\
\label{126}
&=&\beta_{1V}\eta_{2H}+\beta_{2H}\eta_{1V}\\
\label{127}
&=&\beta_{1V}\eta_{2V}+\beta_{2V}\eta_{1V}\\
\label{128}
&=&\gamma_{1H}\alpha_{2V}+\gamma_{2V}\alpha_{1H}\\
\label{129}
&=&\gamma_{1V}\alpha_{2H}+\gamma_{2H}\alpha_{1V}\\
\label{130}
&=&\gamma_{1V}\alpha_{2V}+\gamma_{2V}\alpha_{1V}\\
\nonumber
&=&\gamma_{1H}\eta_{2H}+\gamma_{2H}\eta_{1H}\\
\nonumber
&=&\gamma_{1H}\eta_{2V}+\gamma_{2V}\eta_{1H}\\
\nonumber
&=&\gamma_{1V}\eta_{2H}+\gamma_{2H}\eta_{1V}\\
\nonumber
&=&\gamma_{1V}\eta_{2V}+\gamma_{2V}\eta_{1V}.
\end{eqnarray}
From Eqs.~(\ref{122}), (\ref{128}), (\ref{129}), and (\ref{130}),
\begin{widetext}
\begin{eqnarray*}
0&=&(\gamma_{1V}\alpha_{2H}+\gamma_{2H}\alpha_{1V})\alpha_{1H}\alpha_{2V}+(\gamma_{1H}\alpha_{2V}+\gamma_{2V}\alpha_{1H})\alpha_{2H}\alpha_{1V}\\
&=&(\gamma_{1H}\alpha_{2H}+\gamma_{2H}\alpha_{1H})\alpha_{1V}\alpha_{2V}+(\gamma_{1V}\alpha_{2V}+\gamma_{2V}\alpha_{1V})\alpha_{1H}\alpha_{2H}\\
&=&\sqrt{p_{\rm suc}}\alpha_{1V}\alpha_{2V}.
\end{eqnarray*}
\end{widetext}
 Accordingly, $\alpha_{1V}\alpha_{2V}=0$ when $p_{\rm suc}>0$ is satisfied. In a case where $\alpha_{1V}=0$, from Eqs.~(\ref{120}) and (\ref{124}), $\alpha_{2V}=0$. On the other hand, in another case where $\alpha_{2V}=0$, from Eqs.~(\ref{119}) and (\ref{124}), $\alpha_{1V}=0$. Thus, $\alpha_{1V}=\alpha_{2V}=0$. Then, we derive $\sqrt{p_{\rm suc}}=\beta_{2V}\alpha_{1H}=\beta_{1V}\alpha_{2H}$ from Eqs.~(\ref{119}) and (\ref{120}). This implies that in order to satisfy that $p_{\rm suc}>0$, $\beta_{1V}\beta_{2V}\neq0$ is required. While, from Eqs.~(\ref{121}), (\ref{125}), (\ref{126}), and (\ref{127}), we obtain the following relation:
\begin{widetext}
\begin{eqnarray*}
0&=&(\beta_{1V}\eta_{2H}+\beta_{2H}\eta_{1V})\beta_{1H}\beta_{2V}+(\beta_{1H}\eta_{2V}+\beta_{2V}\eta_{1H})\beta_{2H}\beta_{1V}\\
&=&(\beta_{1H}\eta_{2H}+\beta_{2H}\eta_{1H})\beta_{1V}\beta_{2V}+(\beta_{1V}\eta_{2V}+\beta_{2V}\eta_{1V})\beta_{1H}\beta_{2H}\\
&=&\sqrt{p_{\rm suc}}\beta_{1V}\beta_{2V}.
\end{eqnarray*}
\end{widetext}
Then $\beta_{1V}\beta_{2V}$=0. As a result, $p_{\rm suc}=0$.

\subsection{$|t\rangle=|D_4^{(2)}\rangle$}
The state conversion from $|\Phi^+\rangle_{1,2}|\Phi^+\rangle_{3,4}$ to the state $|D_4^{(2)}\rangle$ is written as
\begin{eqnarray*}
\Pi_{\rm post}V'|\Phi^+\rangle_{1,2}|\Phi^+\rangle_{3,4}=\sqrt{p_{\rm suc}}|D_4^{(2)}\rangle.
\end{eqnarray*}
We obtain the following relations with inner products of $|s_1s_2\rangle_{1,4}$,
\begin{eqnarray}
\label{87_2}
\cfrac{\Pi_{\rm post}V'|HH\rangle_{2,3}}{2}&=&\cfrac{\sqrt{p_{\rm suc}}|VV\rangle_{2,3}}{\sqrt{6}},\\
\label{88_2}
\cfrac{\Pi_{\rm post}V'|HV\rangle_{2,3}}{2}&=&\cfrac{\sqrt{p_{\rm suc}}(|HV\rangle_{2,3}+|VH\rangle_{2,3})}{\sqrt{6}},\ \ \\
\label{89_2}
\cfrac{\Pi_{\rm post}V'|VH\rangle_{2,3}}{2}&=&\cfrac{\sqrt{p_{\rm suc}}(|HV\rangle_{2,3}+|VH\rangle_{2,3})}{\sqrt{6}},
\end{eqnarray}
and
\begin{eqnarray}
\label{90_2}
\cfrac{\Pi_{\rm post}V'|VV\rangle_{2,3}}{2}=\cfrac{\sqrt{p_{\rm suc}}|HH\rangle_{2,3}}{\sqrt{6}}.
\end{eqnarray}
From inner products of $|s_1s_2\rangle_{2,3}$ using Eqs.~(\ref{87_2}), (\ref{88_2}), (\ref{89_2}), and (\ref{90_2}), we obtain the following relations:
\begin{eqnarray}
\label{66_2}
\sqrt{\cfrac{p_{\rm suc}}{6}}&=&\cfrac{\beta_{1V}\alpha_{2V}+\beta_{2V}\alpha_{1V}}{2}\\
\label{67_2}
&=&\cfrac{\beta_{1H}\eta_{2V}+\beta_{2V}\eta_{1H}}{2}\\
\label{68_2}
&=&\cfrac{\beta_{1V}\eta_{2H}+\beta_{2H}\eta_{1V}}{2}\\
\label{69_2}
&=&\cfrac{\gamma_{1H}\alpha_{2V}+\gamma_{2V}\alpha_{1H}}{2}\\
\label{70_2}
&=&\cfrac{\gamma_{1V}\alpha_{2H}+\gamma_{2H}\alpha_{1V}}{2}\\
\label{71_2}
&=&\cfrac{\gamma_{1H}\eta_{2H}+\gamma_{2H}\eta_{1H}}{2},
\end{eqnarray}
and
\begin{eqnarray}
\label{72_2}
0&=&\beta_{1H}\alpha_{2H}+\beta_{2H}\alpha_{1H}\\
\label{73_2}
&=&\beta_{1V}\alpha_{2H}+\beta_{2H}\alpha_{1V}\\
\label{74_2}
&=&\beta_{1H}\alpha_{2V}+\beta_{2V}\alpha_{1H}\\
\nonumber
&=&\beta_{1H}\eta_{2H}+\beta_{2H}\eta_{1H}\\
\nonumber
&=&\beta_{1V}\eta_{2V}+\beta_{2V}\eta_{1V}\\
\label{77_2}
&=&\gamma_{1H}\alpha_{2H}+\gamma_{2H}\alpha_{1H}\\
\nonumber
&=&\gamma_{1V}\alpha_{2V}+\gamma_{2V}\alpha_{1V}\\
\nonumber
&=&\gamma_{1H}\eta_{2V}+\gamma_{2V}\eta_{1H}\\
\nonumber
&=&\gamma_{1V}\eta_{2H}+\gamma_{2H}\eta_{1V}\\
\nonumber
&=&\gamma_{1V}\eta_{2V}+\gamma_{2V}\eta_{1V}.
\end{eqnarray}
From Eqs.~(\ref{66_2}), (\ref{72_2}), (\ref{73_2}), and (\ref{74_2}),
\begin{widetext}
\begin{eqnarray*}
0&=&(\beta_{1V}\alpha_{2H}+\beta_{2H}\alpha_{1V})\alpha_{1H}\alpha_{2V}+(\beta_{2V}\alpha_{1H}+\beta_{1H}\alpha_{2V})\alpha_{2H}\alpha_{1V}\\
&=&(\beta_{1V}\alpha_{2V}+\beta_{2V}\alpha_{1V})\alpha_{1H}\alpha_{2H}+(\beta_{2H}\alpha_{1H}+\beta_{1H}\alpha_{2H})\alpha_{1V}\alpha_{2V}\\
&=&\sqrt{\cfrac{2p_{\rm suc}}{3}}\alpha_{1H}\alpha_{2H}.
\end{eqnarray*}
\end{widetext}
With $p_{\rm suc}>0$, $\alpha_{1H}\alpha_{2H}=0$. Here, we consider two cases, i.e., (I) $\alpha_{1H}=0$ and (II) $\alpha_{2H}=0$.
\begin{enumerate}
\item[(I)] $\alpha_{1H}=0$\\
From Eq.~(\ref{69_2}), $\alpha_{2V}\gamma_{1H}\neq0$. Then, with Eqs.~(\ref{74_2}) and (\ref{77_2}), $\beta_{1H}=\alpha_{2H}=0$. From Eqs.~(\ref{70_2}) and (\ref{73_2}), $\beta_{2H}=0$ and $\alpha_{1V}\neq0$ are also derived.
\item[(II)] $\alpha_{2H}=0$\\
From Eq.~(\ref{70_2}), $\alpha_{1V}\gamma_{2H}\neq0$. Then, with Eqs.~(\ref{73_2}) and (\ref{77_2}), $\beta_{2H}=\alpha_{1H}=0$. From Eqs.~(\ref{69_2}) and (\ref{74_2}), $\beta_{1H}=0$ and $\alpha_{2V}\neq0$ are also derived.
\end{enumerate}
From above results, $\alpha_{1H}=\alpha_{2H}=\beta_{1H}=\beta_{2H}=0$ and $\alpha_{1V}\alpha_{2V}\neq0$ are given. Then, from Eqs.~(\ref{66_2})--(\ref{71_2}),
\begin{eqnarray}
\nonumber
\sqrt{\cfrac{2p_{\rm suc}}{3}}&=&\gamma_{1H}\eta_{2H}+\gamma_{2H}\eta_{1H}\\
\nonumber
&=&\cfrac{2p_{\rm suc}}{3}\left(\cfrac{1}{\alpha_{2V}\beta_{1V}}+\cfrac{1}{\alpha_{1V}\beta_{2V}}\right)\ \\
\nonumber
&=&\cfrac{2p_{\rm suc}}{3}\cfrac{\alpha_{1V}\beta_{2V}+\alpha_{2V}\beta_{1V}}{\alpha_{1V}\beta_{2V}\alpha_{2V}\beta_{1V}}\\
\label{113}
\alpha_{1V}\beta_{2V}\alpha_{2V}\beta_{1V}&=&\cfrac{2p_{\rm suc}}{3}.
\end{eqnarray}
From Eqs.~(\ref{66_2}) and (\ref{113}), we obtain
\begin{eqnarray}
\nonumber
\cfrac{2p_{\rm suc}}{3}&=&\alpha_{1V}\beta_{2V}\alpha_{2V}\beta_{1V}\\
\nonumber
&=&\left(\sqrt{\cfrac{2p_{\rm suc}}{3}}-\beta_{2V}\alpha_{1V}\right)\beta_{2V}\alpha_{1V}\ \ \ \ \\
\label{116_2}
e^{\pm i\pi/3}\sqrt{\cfrac{2p_{\rm suc}}{3}}&=&\beta_{2V}\alpha_{1V}\\
\nonumber
&=&\sqrt{\cfrac{2p_{\rm suc}}{3}}-\beta_{1V}\alpha_{2V}\\
\label{118_2}
\beta_{1V}\alpha_{2V}&=&e^{\mp i\pi/3}\sqrt{\cfrac{2p_{\rm suc}}{3}}
\end{eqnarray}
and then
\begin{widetext}
\begin{eqnarray*}
\cfrac{2p_{\rm suc}}{3}=|\beta_{2V}\alpha_{1V}|^2=|\beta_{1V}\alpha_{2V}|^2=|\eta_{1H}\beta_{2V}|^2=|\eta_{2H}\beta_{1V}|^2=|\gamma_{1H}\alpha_{2V}|^2=|\gamma_{2H}\alpha_{1V}|^2.
\end{eqnarray*}
\end{widetext}
Accordingly, $p_{\rm suc}$ is maximized when $|\alpha_{1V}|^2+|\alpha_{2V}|^2=1$ is satisfied. With $[V'a_{1H}^\dag V'^\dag,V'a_{2H}^\dag V'^\dag]=0$, and Eqs.~(\ref{116_2}) and (\ref{118_2}), we get the following relation:
\begin{eqnarray*}
0&=&\alpha_{1V}^\ast\beta_{1V}+\alpha_{2V}^\ast\beta_{2V}\\
&=&\alpha_{1V}^\ast\cfrac{e^{\mp i\pi/3}}{\alpha_{2V}}\sqrt{\cfrac{2p_{\rm suc}}{3}}+\alpha_{2V}^\ast\cfrac{e^{\pm i\pi/3}}{\alpha_{1V}}\sqrt{\cfrac{2p_{\rm suc}}{3}}\\
&=&\sqrt{\cfrac{p_{\rm suc}}{6}}[1\mp i\sqrt{3}(|\alpha_{1V}|^2-|\alpha_{2V}|^2)].
\end{eqnarray*}
As a result, $p_{\rm suc}=0$.

\subsection{$|t\rangle=|\Phi^+\rangle_{1,3}|\Phi^+\rangle_{2,4}$}
 $p_{\rm suc}=1$ can be achieved because a swap operation between photons in nodes $2$ and $3$ can be done using passive linear optics with unit probability.

\subsection{$|t\rangle=|\Phi^+\rangle_{1,4}|\Phi^+\rangle_{2,3}$}
The state conversion from $|\Phi^+\rangle_{1,2}|\Phi^+\rangle_{3,4}$ to  $|\Phi^+\rangle_{1,4}|\Phi^+\rangle_{2,3}$ is written by
\begin{eqnarray*}
\Pi_{\rm post}V'|\Phi^+\rangle_{1,2}|\Phi^+\rangle_{3,4}=\sqrt{p_{\rm suc}}|\Phi^+\rangle_{1,4}|\Phi^+\rangle_{2,3}.
\end{eqnarray*}
By taking inner product of $|HH\rangle_{1,4}$, we obtain the following relation:
\begin{eqnarray}
\label{124_2}
\Pi_{\rm post}V'|HH\rangle_{2,3}&=&\sqrt{p_{\rm suc}}(|HH\rangle_{2,3}+|VV\rangle_{2,3}).\ \
\end{eqnarray}
By also taking inner products of $|s_1s_1\rangle_{2,3}$ with Eq.~(\ref{124_2}), we obtain the following equation:
\begin{eqnarray*}
\sqrt{p_{\rm suc}}=\beta_{1H}\alpha_{2H}+\beta_{2H}\alpha_{1H}=\beta_{1H}\alpha_{2V}+\beta_{2V}\alpha_{1V}
\end{eqnarray*}
and then
\begin{eqnarray}
\nonumber
\sqrt{p_{\rm suc}}&=&\beta_{1H}\alpha_{2H}+\beta_{2H}\alpha_{1H}\\
\nonumber
&=&\cfrac{\beta_{1H}\alpha_{2H}+\beta_{2H}\alpha_{1H}+\beta_{1H}\alpha_{2V}+\beta_{2V}\alpha_{1V}}{2}\\
\nonumber
&=&\left|\cfrac{\beta_{1H}\alpha_{2H}+\beta_{2H}\alpha_{1H}+\beta_{1H}\alpha_{2V}+\beta_{2V}\alpha_{1V}}{2}\right|\\
\label{165}
&\le&\cfrac{|\beta_{1H}\alpha_{2H}|+|\beta_{2H}\alpha_{1H}|+|\beta_{1H}\alpha_{2V}|+|\beta_{2V}\alpha_{1V}|}{2}.\ \ \ \ \ \ \ \ \
\end{eqnarray}
Equation~(\ref{165}) is maximized when $\sum_{j=1}^2(|\alpha_{jH}|^2+|\alpha_{jV}|^2)=\sum_{j=1}^2(|\beta_{jH}|^2+|\beta_{jV}|^2)=1$ is satisfied. In order to derive the maximum value of $p_{\rm suc}$, we use the method of Lagrange multiplier. First, using variables $\lambda$ and $\mu$, we define a function $f$ as
\begin{widetext}
\begin{eqnarray*}
f\equiv\cfrac{|\beta_{1H}\alpha_{2H}|+|\beta_{2H}\alpha_{1H}|+|\beta_{1H}\alpha_{2V}|+|\beta_{2V}\alpha_{1V}|}{2}+\lambda\left[\sum_{j=1}^2(|\beta_{jH}|^2+|\beta_{jV}|^2)-1\right]+\mu\left[\sum_{j=1}^2(|\alpha_{jH}|^2+|\alpha_{jV}|^2)-1\right].\ \ \ \
\end{eqnarray*}
\end{widetext}
Second, we calculate
\begin{eqnarray}
\label{167}
\cfrac{\partial f}{\partial x}=0,
\end{eqnarray}
where $x\in\{|\alpha_{1H}|,|\alpha_{1V}|,|\alpha_{2H}|,|\alpha_{2V}|,|\beta_{1H}|,|\beta_{1V}|,|\beta_{2H}|,$\\$|\beta_{2V}|,\lambda,\mu\}$. From Eq.~(\ref{167}), we obtain the following relations:
\begin{eqnarray}
\label{132_2}
|\alpha_{2H}|&=&-4\lambda|\beta_{1H}|,
\end{eqnarray}
\begin{eqnarray}
|\alpha_{2V}|&=&-4\lambda|\beta_{1V}|,
\end{eqnarray}
\begin{eqnarray}
|\alpha_{1H}|&=&-4\lambda|\beta_{2H}|,
\end{eqnarray}
\begin{eqnarray}
|\alpha_{1V}|&=&-4\lambda|\beta_{2V}|,
\end{eqnarray}
\begin{eqnarray}
\sum_{j=1}^2(|\alpha_{jH}|^2+|\alpha_{jV}|^2)&=&1,
\end{eqnarray}
and
\begin{eqnarray}
\label{137_2}
\sum_{j=1}^2(|\beta_{jH}|^2+|\beta_{jV}|^2)&=&1.
\end{eqnarray}
From Eqs.~(\ref{132_2})--(\ref{137_2}), $\lambda=-1/4$. Then, we obtain $\sqrt{p_{\rm suc}}\le1/2$. Note that $\lambda=-1/4$ gives the (local) maximum of Eq.~(\ref{165}). As the result, $p_{\rm suc}\le 1/4$.

\subsection{$|t\rangle=|\chi\rangle$}
The state conversion  to $|\chi\rangle$ from $|\Phi^+\rangle_{1,2}|\Phi^+\rangle_{3,4}$ is given as
\begin{eqnarray*}
\Pi_{\rm post}V'|\Phi^+\rangle_{1,2}|\Phi^+\rangle_{3,4}=\sqrt{p_{\rm suc}}|\chi\rangle.
\end{eqnarray*}
We take the following relations from inner products of $|s_1s_2\rangle_{1,4}$:
\begin{eqnarray}
\label{139_2}
\Pi_{\rm post}V'|HH\rangle_{2,3}&=&\sqrt{\cfrac{p_{\rm suc}}{2}}(|HH\rangle_{2,3}+|VV\rangle_{2,3}),\\
\label{140_2}
\Pi_{\rm post}V'|HV\rangle_{2,3}&=&-\sqrt{\cfrac{p_{\rm suc}}{2}}(|HV\rangle_{2,3}+|VH\rangle_{2,3}),\ \ \ \ \ \ \ \\
\label{141_2}
\Pi_{\rm post}V'|VH\rangle_{2,3}&=&\sqrt{\cfrac{p_{\rm suc}}{2}}(|HV\rangle_{2,3}+|VH\rangle_{2,3}),
\end{eqnarray}
and
\begin{eqnarray}
\label{142_2}
\Pi_{\rm post}V'|VV\rangle_{2,3}=\sqrt{\cfrac{p_{\rm suc}}{2}}(|HH\rangle_{2,3}+|VV\rangle_{2,3}).
\end{eqnarray}
From inner products of $|s_1s_2\rangle_{2,3}$ with Eqs.~(\ref{139_2}), (\ref{140_2}), (\ref{141_2}), and (\ref{142_2}), we obtain the following equations:
\begin{eqnarray}
\label{143_2}
\sqrt{\cfrac{p_{\rm suc}}{2}}&=&\beta_{1H}\alpha_{2H}+\beta_{2H}\alpha_{1H}\\
\label{144_2}
&=&\beta_{1V}\alpha_{2V}+\beta_{2V}\alpha_{1V}\\
\label{145_2}
&=&-(\beta_{1H}\eta_{2V}+\beta_{2V}\eta_{1H})\\
\nonumber
&=&-(\beta_{1V}\eta_{2H}+\beta_{2H}\eta_{1V})\\
\label{147_2}
&=&\gamma_{1H}\alpha_{2V}+\gamma_{2V}\alpha_{1H}\\
\nonumber
&=&\gamma_{1V}\alpha_{2H}+\gamma_{2H}\alpha_{1V}\\
\label{149_2}
&=&\gamma_{1H}\eta_{2H}+\gamma_{2H}\eta_{1H}\\
\nonumber
&=&\gamma_{1V}\eta_{2V}+\gamma_{2V}\eta_{1V},
\end{eqnarray}
and
\begin{eqnarray}
\label{151_2}
0&=&\beta_{1H}\alpha_{2V}+\beta_{2V}\alpha_{1H}\\
\label{152_2}
&=&\beta_{1V}\alpha_{2H}+\beta_{2H}\alpha_{1V}\\
\label{153_2}
&=&\beta_{1H}\eta_{2H}+\beta_{2H}\eta_{1H}\\
\nonumber
&=&\beta_{1V}\eta_{2V}+\beta_{2V}\eta_{1V}\\
\label{155_2}
&=&\gamma_{1H}\alpha_{2H}+\gamma_{2H}\alpha_{1H}\\
\nonumber
&=&\gamma_{1V}\alpha_{2V}+\gamma_{2V}\alpha_{1V}\\
\label{157_2}
&=&\gamma_{1H}\eta_{2V}+\gamma_{2V}\eta_{1H}\\
\nonumber
&=&\gamma_{1V}\eta_{2H}+\gamma_{2H}\eta_{1V}.
\end{eqnarray}
When $\alpha_{is}=0$ $(i\in\{1,2\},s\in\{H,V\})$ is satisfied,
\begin{eqnarray*}
\sqrt{\cfrac{p_{\rm suc}}{2}}=\beta_{is}\alpha_{\overline{i}s}=\gamma_{is}\alpha_{\overline{i}\overline
{s}},
\end{eqnarray*}
and
\begin{eqnarray*}
0&=&\beta_{is}\alpha_{\overline{i}\overline{s}}=\gamma_{is}\alpha_{\overline{i}s},
\end{eqnarray*}
where $\overline{1}\equiv2$, $\overline{2}\equiv1$, $\overline{H}\equiv V$, and $\overline{V}\equiv H$. Finally, we get $p_{\rm suc}=0$. With the same way of the above process, we also calculate with $\beta_{i's'}$, $\gamma_{\tilde{i}\tilde{s}}$, and $\eta_{\tilde{i'}\tilde{s'}}$ ($i',\tilde{i},\tilde{i'}\in\{1,2\}$, $s',\tilde{s},\tilde{s'}\in\{H,V\}$) and then $\alpha_{is}\beta_{i's'}\gamma_{\tilde{i}\tilde{s}}\eta_{\tilde{i'}\tilde{s'}}\neq0$, when $p_{\rm suc}>0$. From Eqs.~(\ref{143_2}), (\ref{144_2}), (\ref{151_2}), and (\ref{152_2}),
\begin{widetext}
\begin{eqnarray}
\label{161_2}
0&=&(\beta_{1H}\alpha_{2V}+\beta_{2V}\alpha_{1H})\alpha_{1V}\alpha_{2H}+(\beta_{1V}\alpha_{2H}+\beta_{2H}\alpha_{1V})\alpha_{1H}\alpha_{2V}\\
\nonumber
&=&(\beta_{1H}\alpha_{2H}+\beta_{2H}\alpha_{1H})\alpha_{1V}\alpha_{2V}+(\beta_{1V}\alpha_{2V}+\beta_{2V}\alpha_{1V})\alpha_{1H}\alpha_{2H}\\
\nonumber
&=&\sqrt{\cfrac{p_{\rm suc}}{2}}(\alpha_{1V}\alpha_{2V}+\alpha_{1H}\alpha_{2H}).
\end{eqnarray}
\end{widetext}
Replacing $\alpha$ with $\beta$ in Eq.~(\ref{161_2}),
\begin{eqnarray*}
0=\sqrt{\cfrac{p_{\rm suc}}{2}}(\beta_{1V}\beta_{2V}+\beta_{1H}\beta_{2H}).
\end{eqnarray*}
With the same process for other equations and $p_{\rm suc}>0$,
\begin{eqnarray}
\label{165_2}
\alpha_{1V}\alpha_{2V}+\alpha_{1H}\alpha_{2H}&=&0,\\
\nonumber
\beta_{1V}\beta_{2V}+\beta_{1H}\beta_{2H}&=&0,\\
\label{167_2}
\alpha_{1V}\alpha_{2H}+\alpha_{1H}\alpha_{2V}&=&0,\\
\nonumber
\beta_{1V}\beta_{2H}+\beta_{1H}\beta_{2V}&=&0,\\
\nonumber
\gamma_{1V}\gamma_{2V}+\gamma_{1H}\gamma_{2H}&=&0,\\
\nonumber
\eta_{1V}\eta_{2V}+\eta_{1H}\eta_{2H}&=&0,\\
\nonumber
\gamma_{1V}\gamma_{2H}+\gamma_{1H}\gamma_{2V}&=&0,
\end{eqnarray}
and
\begin{eqnarray*}
\eta_{1V}\eta_{2H}+\eta_{1H}\eta_{2V}&=&0.
\end{eqnarray*}
From Eqs.~(\ref{165_2}) and (\ref{167_2}),
\begin{eqnarray*}
-\cfrac{\alpha_{1H}\alpha_{2V}}{\alpha_{2H}}&=&-\cfrac{\alpha_{1H}\alpha_{2H}}{\alpha_{2V}}\\
\cfrac{\alpha_{2V}}{\alpha_{2H}}&=&\cfrac{\alpha_{2H}}{\alpha_{2V}}\\
\alpha_{2V}&=&(-1)^a\alpha_{2H}.
\end{eqnarray*}
Here, $a\in\{0,1\}$. From the same equations, we also obtain the following equation,
\begin{eqnarray*}
-\cfrac{\alpha_{1V}\alpha_{2V}}{\alpha_{1H}}&=&-\cfrac{\alpha_{1H}\alpha_{2V}}{\alpha_{1V}}\\
\cfrac{\alpha_{1V}}{\alpha_{1H}}&=&\cfrac{\alpha_{1H}}{\alpha_{1V}}\\
\alpha_{1V}&=&-(-1)^a\alpha_{1H}.
\end{eqnarray*}
By using the same way for other equations, $\beta_{2V}=(-1)^b\beta_{2H}$, $\beta_{1V}=-(-1)^b\beta_{1H}$, $\gamma_{2V}=(-1)^c\gamma_{2H}$, $\gamma_{1V}=-(-1)^c\gamma_{1H}$, $\eta_{2V}=(-1)^d\eta_{2H}$, and $\eta_{1V}=-(-1)^d\eta_{1H}$ are also derived. Here, $b,c,d\in\{0,1\}$. From Eqs.~(\ref{143_2}) and (\ref{151_2}),
\begin{eqnarray*}
\sqrt{\cfrac{p_{\rm suc}}{2}}&=&\beta_{1H}\alpha_{2H}+\beta_{2H}\alpha_{1H}\\
&=&(-1)^a\beta_{1H}\alpha_{2V}+(-1)^b\beta_{2V}\alpha_{1H}\\
&=&[(-1)^a-(-1)^b]\beta_{1H}\alpha_{2V}.
\end{eqnarray*}
Here, $a=b\oplus 1$ has to be held to satisfy $p_{\rm suc}>0$. From Eqs.~(\ref{145_2}) and (\ref{153_2}),
\begin{eqnarray*}
\sqrt{\cfrac{p_{\rm suc}}{2}}&=&-(\beta_{1H}\eta_{2V}+\beta_{2V}\eta_{1H})\\
&=&-[(-1)^d\beta_{1H}\eta_{2H}+(-1)^b\beta_{2H}\eta_{1H}]\\
&=&-[(-1)^d-(-1)^b]\beta_{1H}\eta_{2H}.
\end{eqnarray*}
Then, $b=d\oplus 1$ holds with $p_{\rm suc}>0$. From Eqs.~(\ref{147_2}) and (\ref{155_2}),
\begin{eqnarray*}
\sqrt{\cfrac{p_{\rm suc}}{2}}&=&\gamma_{1H}\alpha_{2V}+\gamma_{2V}\alpha_{1H}\\
&=&(-1)^a\gamma_{1H}\alpha_{2H}+(-1)^c\gamma_{2H}\alpha_{1H}\\
&=&[(-1)^a-(-1)^c]\gamma_{1H}\alpha_{2H}\\
&=&[1-(-1)^{a\oplus c}]\gamma_{1H}\alpha_{2V},
\end{eqnarray*}
and then $a=c\oplus 1$, when $p_{\rm suc}>0$. From Eqs.~(\ref{149_2}) and (\ref{157_2}),
\begin{eqnarray*}
\sqrt{\cfrac{p_{\rm suc}}{2}}&=&\gamma_{1H}\eta_{2H}+\gamma_{2H}\eta_{1H}\\
&=&(-1)^d\gamma_{1H}\eta_{2V}+(-1)^c\gamma_{2V}\eta_{1H}\\
&=&[(-1)^d-(-1)^c]\gamma_{1H}\eta_{2V}\\
&=&[1-(-1)^{c\oplus d}]\gamma_{1H}\eta_{2H},
\end{eqnarray*}
and then $c=d\oplus 1$, when $p_{\rm suc}>0$. Thus, 
\begin{widetext}
\begin{eqnarray*}
\sqrt{\cfrac{p_{\rm suc}}{2}}=(-1)^a2\beta_{1H}\alpha_{2V}=-(-1)^a2\beta_{1H}\eta_{2H}=2\gamma_{1H}\alpha_{2V}=2\gamma_{1H}\eta_{2H},
\end{eqnarray*}
\end{widetext}
and then $\alpha_{2V}=-\eta_{2H}=-\alpha_{2V}$. However, this contradicts $\alpha_{is}\beta_{i's'}\gamma_{\tilde{i}\tilde{s}}\eta_{\tilde{i'}\tilde{s'}}\neq0$. Accordingly, $p_{\rm suc}=0$.

\medskip
\section*{\red APPENDIX B}
{\red In this Appendix, we derive Eqs.~(\ref{p_suc_im}) and (\ref{fidelity_im}). We define $U_{\rm PDBS}$ as a unitary operator denoting the function of the PDBS with transmittance $T_H$ and $T_V$. By using this definition and Pauli operators $X_i$ and $Z_i$ acting on spatial mode $i$, the success probability can be written as
\begin{widetext}
\begin{eqnarray*}
&&p_{\rm suc}\\
&=&\eta\eta'||\Pi_{\rm post}Z_{2'}X_{2'}X_{1'}U_{\rm PDBS}X_2X_1U_{\rm PDBS}U_{\rm PDBS}|\Phi^+\rangle|\Phi^+\rangle||^2\\
&=&\eta\eta'\left|\left|\Pi_{\rm post}Z_{2'}X_{2'}X_{1'}U_{\rm PDBS}X_2X_1\cfrac{T_H|HHHH\rangle+\sqrt{T_HT_V}(|HHVV\rangle+|VVHH\rangle)+T_V|VVVV\rangle}{2}\right|\right|^2\\
&=&\eta\eta'\left|\left|\Pi_{\rm post}Z_{2'}X_{2'}X_{1'}U_{\rm PDBS}\cfrac{T_H|HVVH\rangle+\sqrt{T_HT_V}(|HVHV\rangle+|VHVH\rangle)+T_V|VHHV\rangle}{2}\right|\right|^2\\
&=&\cfrac{\eta\eta'}{4}\large|\large|T_H(1-2T_V)|HHHH\rangle+T_HT_V(|HHVV\rangle+|VVHH\rangle)\\
&&+\sqrt{T_H(1-T_H)T_V(1-T_V)}(|HVHV\rangle+|VHVH\rangle)-T_V(2T_H-1)|VVVV\rangle\large|\large|^2\\
&=&\cfrac{\eta\eta'}{4}[T_H^2(1-2T_V)^2+2T_H^2T_V^2+2T_H(1-T_H)T_V(1-T_V)+T_V^2(2T_H-1)^2]\\
&=&\cfrac{\eta\eta'(T_H^2+2T_HT_V+T_V^2-6T_H^2T_V-6T_HT_V^2+12T_H^2T_V^2)}{4},
\end{eqnarray*}
\end{widetext}
where $|||\psi\rangle||\equiv\sqrt{\langle\psi|\psi\rangle}$. From above calculation, when the postselection succeeds, the output state $|\psi_{\rm out}\rangle$ is
\begin{eqnarray*}
|\psi_{\rm out}\rangle&=&\cfrac{\sqrt{\eta\eta'}}{2\sqrt{p_{\rm suc}}}[T_H(1-2T_V)|HHHH\rangle\\
&&+T_HT_V(|HHVV\rangle+|VVHH\rangle)\\
&&+\sqrt{T_H(1-T_H)T_V(1-T_V)}(|HVHV\rangle+|VHVH\rangle)\\
&&-T_V(2T_H-1)|VVVV\rangle].
\end{eqnarray*}
Accordingly, the fidelity $F$ is
\begin{eqnarray*}
F&=&\sqrt{|\langle C_4|\psi_{\rm out}\rangle|^2}\\
&=&|\langle C_4|\psi_{\rm out}\rangle|\\
&=&\cfrac{|T_H(1-2T_V)+2T_HT_V+T_V(2T_H-1)|}{2\sqrt{T_H^2+2T_HT_V+T_V^2-6T_H^2T_V-6T_HT_V^2+12T_H^2T_V^2}}\\
&=&\cfrac{|T_H+2T_HT_V-T_V|}{2\sqrt{T_H^2+2T_HT_V+T_V^2-6T_H^2T_V-6T_HT_V^2+12T_H^2T_V^2}}.
\end{eqnarray*}}


\begin{thebibliography}{99}
\bibitem{[E91]}A. K. Ekert, Quantum Cryptography Based on Bell's Theorem, Phys. Rev. Lett. {\bf 67}, 661 (1991).
\bibitem{[BBCJPW93]}C. H. Bennett, G. Brassard, C. Cr\'{e}peau, R. Jozsa, A. Peres, and W. K. Wootters, Teleporting an unknown quantum state via dual classical and Einstein-Podolsky-Rosen channels, Phys. Rev. Lett. {\bf 70}, 1895 (1993).
\bibitem{[YC06]}Y. Yeo and W. K. Chua, Teleportation and Dense Coding with Genuine Multipartite Entanglement, Phys. Rev. Lett. {\bf 96}, 060502 (2006).
\bibitem{[BR03]}H. Buhrman and H. R\"{o}hrig, {\it Distributed Quantum Computing, 28th International Symposium, Mathematical Foundations of Computer Science} (Bratislava Slovakia, 2003), pp. 1--20.
\bibitem{[TKM05]}S. Tani, H. Kobayashi, and K. Matsumoto, Exact quantum algorithms for the leader election problem, Proc. of the 22nd Symposium on Theoretical Aspects of Computer Science (STACS05), LNCS Vol. 3404 (2005).
\bibitem{[DP06]}E. D'Hondt and P. Panangaden, The computational power of the W and GHZ states, Quantum Inf. Comput. {\bf 6}, 173 (2006).
\bibitem{[BR01]}H. J. Briegel and R. Raussendorf, Persistent Entanglement in Arrays of Interacting Particles, Phys. Rev. Lett. {\bf 86}, 910 (2001).
\bibitem{[RB01]}R. Raussendorf and H. J. Briegel, A One-Way Quantum Computer, Phys. Rev. Lett. {\bf 86}, 5188 (2001).
\bibitem{[RBB03]}R. Raussendorf, D. E. Browne, and H. J. Briegel, Measurement-based quantum computation on cluster states, Phys. Rev. A {\bf 68}, 022312 (2003).
\bibitem{[RHBM13]}M. Rossi, M. Huber, D. Bru\ss, and C. Macchiavello, Quantum hypergraph states, New J. Phys. {\bf 15}, 113022 (2013).
\bibitem{[DVC00]}W. D\"{u}r, G. Vidal, and J. I. Cirac, Three qubits can be entangled in two inequivalent ways, Phys. Rev. A {\bf 62}, 062314 (2000).
\bibitem{[KBI00]}M. Koashi, V. Bu\v{z}ek, and N. Imoto, Entangled webs: Tight bound for symmetric sharing of entanglement, Phys. Rev. A {\bf 62}, 050302(R) (2000).
\bibitem{[D01]}W. D\"{u}r, Multipartite entanglement that is robust against disposal of particles, Phys. Rev. A {\bf 63}, 020303(R) (2001).
\bibitem{[GHZ89]}D. M. Greenberger, M. A. Horne, and A. Zeilinger, Going Beyond Bell's Theorem, in {\it Bell's Theorem, Quantum Theory, and Conceptions of the Universe} (Kluwer, Dordrecht, 1989), pp. 69--72.
\bibitem{[BFK09]}A. Broadbent, J. Fitzsimons, and E. Kashefi, Universal Blind Quantum Computation, in {\it Proceedings of the 50th Annual Symposium on Foundations of Computer Science} (IEEE Computer Society, Los Alamitos, 2009), pp. 517--526.
\bibitem{[TFIYI16]}Y. Takeuchi, K. Fujii, R. Ikuta, T. Yamamoto, and N. Imoto, Blind quantum computation over a collective-noise channel, Phys. Rev. A {\bf 93}, 052307 (2016).
\bibitem{[HPF15]}M. Hajdu\v{s}ek, C. A. P\'{e}rez-Delgado, and J. F. Fitzsimons, Device-Independent Verifiable Blind Quantum Computation, arXiv:1502.02563.
\bibitem{[MM16]}J. Miller and A. Miyake, Hierarchy of universal entanglement in 2D measurement-based quantum computation, npj Quantum Information {\bf 2}, 16036 (2016).
\bibitem{[SIGA05]}V. Scarani, S. Iblisdir, N. Gisin, and A. Ac\'\i n, Quantum cloning, Rev. Mod. Phys. {\bf 77}, 1225 (2005).
\bibitem{[DKK16]}V. Dunjko, T. Kapourniotis, and E. Kashefi, Quantum-enhanced secure delegated classical computing, Quantum Inf. Comput. {\bf 16}, 61 (2016).
\bibitem{[D54]}R. H. Dicke, Coherence in Spontaneous Radiation Processes, Phys. Rev. {\bf 93}, 99 (1954).
\bibitem{[IILV10]}S. S. Ivanov, P. A. Ivanov, I. E. Linington, and N. V. Vitanov, Scalable quantum search using trapped ions, Phys. Rev. A {\bf 81}, 042328 (2010).
\bibitem{[SOMI04]}J. Shimamura, \c{S}. K. \"{O}zdemir, F. Morikoshi, and N. Imoto, Entangled states that cannot reproduce original classical games in their quantum version, Phys. Lett. A {\bf 328}, 20 (2004).
\bibitem{[OSI07]}\c{S}. K. \"{O}zdemir, J. Shimamura, and N. Imoto, A necessary and sufficient condition to play games in quantum mechanical settings, New J. Phys. {\bf 9}, 43 (2007).
\bibitem{[WYKO07]}C. Wu, Y. Yeo, L. C. Kwek, and C. H. Oh, Quantum nonlocality of four-qubit entangled states, Phys. Rev. A {\bf 75}, 032332 (2007).
\bibitem{[KSTSW07]}N. Kiesel, C. Schmid, G. T\'{o}th, E. Solano, and H. Weinfurter, Experimental Observation of Four-Photon Entangled Dicke State With High Fidelity, Phys. Rev. Lett. {\bf 98}, 063604 (2007).
\bibitem{[WRZ05]}P. Walther, K. J. Resch, and A. Zeilinger, Local Conversion of Greenberger-Horne-Zeilinger States to Approximate W States, Phys. Rev. Lett. {\bf 94}, 240501 (2005).
\bibitem{[TOYKI08]}T. Tashima, \c{S}. K. \"{O}zdemir, T. Yamamoto, M. Koashi, and N. Imoto, Elementary optical gate for expanding an entanglement web, Phys. Rev. A {\bf 77}, 030302(R) (2008).
\bibitem{[TOYKI09]}T. Tashima, \c{S}. K. \"{O}zdemir, T. Yamamoto, M. Koashi, and N. Imoto, Local expansion of photonic W state using a polarization-dependent beamsplitter, New J. Phys. {\bf 11}, 023024 (2009).
\bibitem{[ITYKI11]}R. Ikuta, T. Tashima, T. Yamamoto, M. Koashi, and N. Imoto, Optimal local expansion of $W$ states using linear optics and Fock states, Phys. Rev. A {\bf 83}, 012314 (2011).
\bibitem{[VDMV02]}F. Verstraete, J. Dehaene, B. De Moor, and H. Verschelde, Four qubits can be entangled in nine different ways, Phys. Rev. A {\bf 65}, 052112 (2002).
\bibitem{[SKLWZW08]}C. Schmid, N. Kiesel, W. Laskowski, W. Wieczorek, M. \.{Z}ukowski, and H. Weinfurter, Discriminating Multipartite Entangled States, Phys. Rev. Lett. {\bf 100}, 200407 (2008).
\bibitem{[TTONKW16]}T. Tashima, M. S. Tame, \c{S}. K. \"{O}zdemir, F. Nori, M. Koashi, and H. Weinfurter, Photonic multipartite entanglement conversion using nonlocal operations, Phys. Rev. A {\bf 94}, 052309 (2016).
\bibitem{[MSMMSJTOT17]}M. Mi\v{c}uda, R. St\'{a}rek, P. Marek, M. Mikov\'{a}, I. Straka, M. Je\v{z}ek, T. Tashima, \c{S}. K. \"{O}zdemir, and M. Tame, Experimental characterization of a non-local convertor for quantum photonic networks, Opt. Express {\bf 25}, 7839 (2017).
\bibitem{[KIOTYKI14]}T. Kobayashi, R. Ikuta, \c{S}. K. \"{O}zdemir, M. Tame, T. Yamamoto, M. Koashi, and N. Imoto, Universal gates for transforming multipartite entangled Dicke states, New J. Phys. {\bf 16}, 023005 (2014).
\bibitem{[OBYATO14]}F. Ozaydin, S. Bugu, C. Yesilyurt, A. A. Altintas, M. Tame, and \c{S}. K. \"{O}zdemir, Fusing multiple W states simultaneously with a Fredkin gate, Phys. Rev. A {\bf 89}, 042311 (2014).
\bibitem{[ZVSW03]}J. Zhang, J. Vala, S. Sastry, and K. Birgitta Whaley, Geometric theory of nonlocal two-qubit operations, Phys. Rev. A {\bf 67}, 042313 (2003).
\bibitem{[SNBMLM06]}T. P. Spiller, K. Nemoto, S. L. Braunstein, W. J. Munro, P. van Loock, and G. J. Milburn, Quantum computation by communication, New J. Phys. {\bf 8}, 30 (2006).
\bibitem{[KSWUW05]}N. Kiesel, C. Schmid, U. Weber, R. Ursin, and H. Weinfurter, Linear Optics Controlled-Phase Gate Made Simple, Phys. Rev. Lett. {\bf 95}, 210505 (2005).
\bibitem{[OHTS05]}R. Okamoto, H. F. Hofmann, S. Takeuchi, and K. Sasaki, Demonstration of an Optical Quantum Controlled-NOT Gate without Path Interference, Phys. Rev. Lett. {\bf 95}, 210506 (2005).
\bibitem{[PDGWZ01]}J.-W. Pan, M. Daniell, S. Gasparoni, G. Weihs, and A. Zeilinger, Experimental Demonstration of Four-Photon Entanglement and High-Fidelity Teleportation, Phys. Rev. Lett. {\bf 86}, 4435 (2001).
\bibitem{[KLM01]}E. Knill, R. Laflamme, and G. Milburn, A scheme for efficient quantum computation with linear optics, Nature (London) {\bf 409}, 46 (2001).
\bibitem{[TWOYKI09]}T. Tashima, T. Wakatsuki, \c{S}. K. \"{O}zdemir, T. Yamamoto, M. Koashi, and N. Imoto, Local transformation of two Einstein-Podolsky-Rosen photon pairs into a three-photon W state, Phys. Rev. Lett. {\bf 102}, 130502 (2009).
\bibitem{[TKYKI08]}Y. Tokunaga, S. Kuwashiro, T. Yamamoto, M. Koashi, and N. Imoto, Generation of High-Fidelity Four-Photon Cluster State and Quantum-Domain Demonstration of One-Way Quantum Computing, Phys. Rev. Lett. {\bf 100}, 210501 (2008).
\bibitem{[WRRSWVAZ05]}P. Walther, K. J. Resch, T. Rudolph, E. Schenck, H. Weinfurter, V. Vedral, M. Aspelmeyer, and A. Zeilinger, Experimental One-Way Quantum Computing, Nature (London) {\bf 434}, 169 (2005).
\end{thebibliography}
\end{document}